\documentclass[a4paper,11pt]{article}
\usepackage[margin=2cm]{geometry}
\usepackage{array}
\usepackage{booktabs}
\usepackage{amsthm}
\usepackage{amsmath}
\usepackage{amstext}
\usepackage{amsfonts}
\usepackage{graphicx}
\usepackage{lineno}
\usepackage{bbm}
\usepackage{xurl}
\usepackage{verbatim}
\usepackage{setspace}
\usepackage[no-weekday]{eukdate}
\usepackage{mathpazo}
\usepackage{bm}

\usepackage{hyperref}
\hypersetup{
  colorlinks=true,
  linkcolor=blue,
  citecolor=blue,
  urlcolor=blue}

\usepackage{verbatim}
\newcommand{\pkg}[1]{{\normalfont\fontseries{b}\selectfont #1}}
\let\proglang=\textsf

\usepackage{todonotes}

\def\citeapos#1{\citeauthor{#1}'s (\citeyear{#1})}

\usepackage[backend=biber, style=authoryear, bibencoding=utf8, minnames=1, maxnames=3, maxcitenames=2, maxbibnames=99, giveninits=true, terseinits=true, firstinits=true, hyperref=true, natbib=true,dashed=false, uniquelist=false, uniquename=false, date=year, doi=true, isbn=false, eprint=false]{biblatex}
\DeclareFieldFormat{url}{\url{#1}}
\DeclareFieldFormat[article]{pages}{#1}
\DeclareFieldFormat[inproceedings]{pages}{\lowercase{pp.}#1}
\DeclareFieldFormat[incollection]{pages}{\lowercase{pp.}#1}
\DeclareFieldFormat[article]{volume}{\mkbibbold{#1}}
\DeclareFieldFormat[article]{number}{\mkbibparens{#1}}
\DeclareFieldFormat[article]{title}{\MakeCapital{#1}}
\DeclareFieldFormat[article]{url}{}
\DeclareFieldFormat[book]{url}{}
\DeclareFieldFormat[inbook]{url}{}
\DeclareFieldFormat[incollection]{url}{}
\DeclareFieldFormat[inproceedings]{url}{}
\DeclareFieldFormat[inproceedings]{title}{#1}
\DeclareFieldFormat{shorthandwidth}{#1}
\DeclareNameAlias{sortname}{family-given}
\AtEveryBibitem{\clearfield{month}}
\AtEveryCitekey{\clearfield{month}}
\renewbibmacro{in:}{%
  \ifentrytype{article}{}{%
  \printtext{\bibstring{in}\intitlepunct}}}
\usepackage{xpatch}
\xpatchbibmacro{volume+number+eid}{\setunit*{\adddot}}{}{}{}
\usepackage{microtype}

\makeatletter
\renewcommand{\paragraph}{\@startsection{paragraph}{4}{0ex}%
   {-3.25ex plus -1ex minus -0.2ex}%
   {1.5ex plus 0.2ex}%
   {\normalfont\normalsize\bfseries}}
\makeatother

\begin{document}

\def\spacingset#1{\renewcommand{\baselinestretch}%
  {#1}\small\normalsize} \spacingset{1.25}

\title{\bf
  Forecast combinations: an over 50-year review
}

\author{Xiaoqian Wang\footnote{School of Economics and Management, Beihang University,
    Beijing 100191, China. E-mail: xiaoqianwang@buaa.edu.cn.},\quad
  Rob J Hyndman\footnote{Department of Econometrics \& Business Statistics, Monash University,
    Clayton VIC 3800, Australia. E-mail: rob.hyndman@monash.edu.},\quad
  Feng Li\footnote{School of Statistics and Mathematics, Central University of Finance and Economics,
    Beijing 102206, China. E-mail: feng.li@cufe.edu.cn.},\quad
  Yanfei Kang\footnote{Author for correspondence. School of Economics and Management, Beihang University,
    Beijing 100191, China. E-mail: yanfeikang@buaa.edu.cn}}

\maketitle

\bigskip
\begin{abstract}
  Forecast combinations have flourished remarkably in the forecasting community and, in recent years, have become part of the mainstream of forecasting research and activities. Combining multiple forecasts produced from single (target) series is now widely used to improve accuracy through the integration of information gleaned from different sources, thereby mitigating the risk of identifying a single ``best'' forecast. Combination schemes have evolved from simple combination methods without estimation, to sophisticated methods involving time-varying weights, nonlinear combinations, correlations among components, and cross-learning. They include combining point forecasts and combining probabilistic forecasts. This paper provides an up-to-date review of the extensive literature on forecast combinations, together with reference to available open-source software implementations. We discuss the potential and limitations of various methods and highlight how these ideas have developed over time. Some important issues concerning the utility of forecast combinations are also surveyed. Finally, we conclude with current research gaps and potential insights for future research.
\end{abstract}

\noindent%
\vfill

\newpage

\spacingset{1.25}

\section{Introduction}
\label{sec:introduction}

The idea of combining multiple individual forecasts dates back at least to Francis Galton, who in 1906 visited an ox-weight-judging competition and observed that the average of 787 estimates of an ox's weight was remarkably close to the ox's actual weight; see \citet{Surowiecki2005-wisdom} for more details of the story. About sixty years later, the famous work of \citet{Bates1969-yj} popularized the idea and spawned a rich literature on forecast combinations. More than fifty years have passed since \citeapos{Bates1969-yj} seminal work, and it is now well established that forecast combinations are beneficial, offering substantially improved forecasts on average relative to constituent models; see \citet{Clemen1989-fb} and \citet{Timmermann2006-en} for extensive earlier literature reviews.

In this paper, we aim to present an up-to-date modern review of the literature on forecast combinations over the past five decades. We cover a wide variety of forecast combination methods for both point forecasts and probabilistic forecasts, contrasting them and highlighting how various related ideas have developed in parallel.

Combining multiple forecasts derived from numerous forecasting methods is often a better approach than identifying a single ``best forecast''. These are usually called ``combination forecasts'' or ``ensemble forecasts'' in different domains. Observed time series data are unlikely to be generated by a simple process specified with a specific functional form because of the possibility of time-varying trends, seasonality changes, structural breaks, and the complexity of real data generating processes \citep{Clements1998-bu}. Thus, selecting a single ``best model'' to approximate the unknown underlying data generating process may be misleading, and is subject to at least three sources of uncertainty: data uncertainty, parameter uncertainty, and model uncertainty \citep{Petropoulos2018-fw,Kourentzes2019-na}. Given these challenges, it is often better to combine multiple forecasts to incorporate multiple drivers of the data generating process and mitigate uncertainties regarding model form and parameter specification.

Potential explanations for the strong performance of forecast combinations are manifold. First, the combination is likely to improve forecasting performance when multiple forecasts to be combined incorporate partial (but incompletely overlapping) information. Second, structural breaks are a common motivation for combining forecasts from different models \citep{Timmermann2006-en}. In the presence of structural breaks and other instabilities, combining forecasts from models with different degrees of misspecification and adaptability can mitigate the problem, and helps explains the empirical success of forecast combinations. See, e.g., \citet{Rossi2013-fi,Rossi2021-fi} for an extensive discussion on forecast combinations in the presence of instabilities. Indeed, one can consider the competing forecasts as a form of intercept correction relative to a baseline forecast, providing potential gains in forecast accuracy if there are either structural breaks or deterministic misspecifications \citep{Hendry2004-pf}. Finally, \citet{Hendry2004-pf} noted that forecast combination can be viewed as an application of Stein-James shrinkage estimation \citep{Judge1978-sr}. Specifically, if the unknown future value is considered as a ``meta-parameter'' of which all the individual forecasts are estimates, then averaging has the potential to provide an improved estimate.

\begin{figure}[!htb]
  \centering \includegraphics[width=0.85\textwidth]{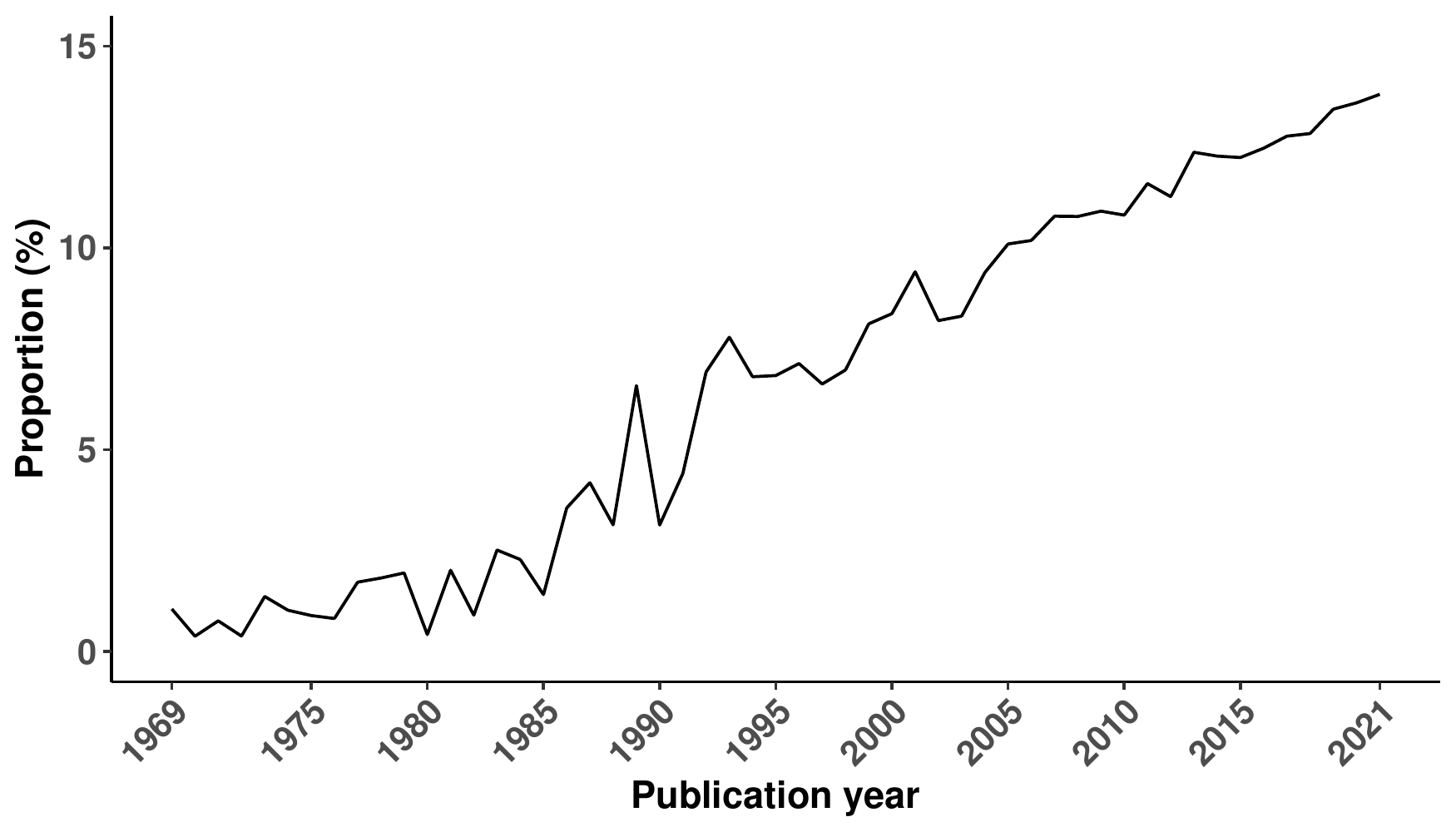}
  \caption{The proportion of papers that concern forecast combinations among all published forecasting papers included in the Web of Science databases during the publication year range 1969--2021. Specifically, we use the search query \texttt{TS = (forecast*)} to find all forecasting papers, and to find papers concerning forecast combinations we use \texttt{TS = ((forecast* NEAR/5 combin*) OR (forecast* NEAR/5 ensemble*) OR (forecast* NEAR/5 averag*) OR (forecast* NEAR/5 aggregat*) OR (forecast* NEAR/5 pool*) OR (forecast* AND ((model* NEAR/5 combin*) OR (model* NEAR/5 ensemble*) OR (model* NEAR/5 averag*) OR (model* NEAR/5 aggregat*) OR (model* NEAR/5 pool*))))}.}
  \label{fig:prop}
\end{figure}

In light of their superiority, forecast combinations have appeared in a wide range of applications such as retail \citep{Ma2021-np}, energy \citep{Xie2016-fb}, economics \citep{Aastveit2019-lf}, and epidemiology \citep{Ray2022-co}. Among all published forecasting papers included in the Web of Science, the proportion of papers concerning forecast combinations has been trending upward over the past $50$ years, reaching $13.80\%$ in $2021$, as shown in Figure~\ref{fig:prop}. As a consequence, it is timely and necessary to review the extant literature on this topic.

The gains from forecast combinations rely on not only the quality of the individual forecasts to be combined, but the estimation of the combination weights assigned to each forecast \citep{Timmermann2006-en,Cang2014-tp}. Numerous studies have been devoted to discussing critical issues concerning the constitution of the model pool and the selection of the optimal model subset, including but not limited to the accuracy, diversity, and robustness of individual models \citep{Batchelor1995-ps,Mannes2014-dl,Thomson2019-al,Lichtendahl2020-ut,Kang2021-ol}. On the other hand, combination schemes vary across studies and have evolved from simple combination methods that avoid weight estimation \citep[e.g.,][]{Clemen1986-pd,Palm1992-im,Genre2013-ut,Grushka-Cockayne2017-dj,Petropoulos2020-fp} to sophisticated methods that tailor weights for different individual models \citep[e.g.,][]{Bates1969-yj,Newbold1974-lp,Kolassa2011-ai,Li2020-od,Montero-Manso2020-tq,Kang2021-ol,Wang2021-un}. Accordingly, forecast combinations can be linear or nonlinear, static or time-varying, series-specific or cross-learning, and ignore or cover correlations among individual forecasts. Despite the diverse set of forecast combination schemes, forecasters still have little guidance on how to solve the ``forecast combination puzzle'' \citep{Stock2004-rq,Smith2009-wd,Claeskens2016-pv,Chan2018-jl} --- simple averaging often empirically dominates sophisticated weighting schemes that should (asymptotically) be superior.

Initial work on forecast combinations after the seminal work of \citet{Bates1969-yj} focused on dealing with point forecasts \citep[see, for example,][]{Clemen1989-fb,Timmermann2006-en}. In recent years considerable attention has moved towards the use of probabilistic forecasts \citep[e.g.,][]{Hall2007-lh,Gneiting2013-hl,Kapetanios2015-bb,Martin2021-yi} as they enable a rich assessment of forecast uncertainties. When working with probabilistic forecasts, issues such as diversity among individual forecasts can be more complex and less understood than combining point forecasts \citep{Ranjan2010-jl}, and additional issues such as calibration and sharpness need to be considered when assessing or selecting a combination scheme \citep{Gneiting2007-fr}. Additionally, probabilistic forecasts can be elicited in different forms (i.e., density forecasts, quantiles, prediction intervals, etc.), and the resulting combinations may have different properties such as calibration, sharpness, and shape; see \citet{Lichtendahl2013-rt} for further analytical details.

We should clarify that we take the individual forecasts to be combined as given, and we do not discuss how the forecasts themselves are generated. We focus our attention on combinations of multiple forecasts derived from \textit{separate} and \textit{non-interfering} models for a given time series. Nevertheless, the literature involves at least two other types of combinations that are not covered in the present review. The first is the case of generating multiple series from the single (target) series, forecasting each of the generated series independently, and then combining the outcomes. Such data manipulation extracts more information from the target time series, which, in turn, can be used to enhance the forecasting performance. \citet{Petropoulos2021-wisdom} referred to this category of forecast combinations generally as ``wisdom of the data'' and provided an overview of approaches in this category. In this particular context, the combination methods reviewed in this paper can function as tools to aggregate (or combine) the forecasts computed from different perspectives of the same data. The second type of forecast combination we do not cover is forecast reconciliation for hierarchical time series, which has developed over the past ten years since the pioneering work of \citet{Hyndman2011-sd}. Forecast reconciliation involves reconciling forecasts across the hierarchy to ensure that the forecasts sum appropriately across the levels of the hierarchy, and hence is a type of forecast combination.

We note that forecast combination and model averaging are sometimes used without distinction in the literature. The two terms overlap, but their focuses are different. ``Model averaging'' is a general term allowing for model uncertainty, particularly in parameter estimation, which can lead to better estimates and more reliable forecasts and prediction intervals than model selection (selecting a single best model). Several approaches to model averaging have been developed in statistics, econometrics, and machine learning. Two main strands can be identified: frequentist approaches \citep[e.g.,][]{Fletcher2018-ma} and Bayesian approaches \citep[e.g.][]{Steel2020-ma}. ``Forecast combination'' is a more focused terminology describing the combination of forecasts to generate a better forecast; the component forecasts could be outcomes from model averaging, individual models, or expert forecasts, for example. As with model averaging, weights can be used to combine the component forecasts. Unlike model averaging, however, forecast combination also has some underlying assumptions on the forecasts to ensure that the forecast combinations are unbiased or optimal.

This paper aims to contribute a broad perspective and historical overview of the main developments in forecast combinations. The paper is organized into two main sections on point forecast combinations (\autoref{sec:point}) and probabilistic forecast combinations (\autoref{sec:probabilistic}). Section~\ref{sec:conclusion} concludes the paper and identifies possible future developments in the future.

\section{Point forecast combinations}
\label{sec:point}

\subsection{Simple point forecast combinations}
\label{sec:simple_comb}

A considerable literature has accumulated over the years regarding how individual forecasts are combined, with the unanimous conclusion that simple combination schemes are hard to beat \citep{Kang1986-kq,Clemen1989-fb,Fischer1999-kz,Stock2004-rq,Lichtendahl2020-ut}. That is, equally weighted averages, which ignore past information regarding the precision of individual forecasts and correlations between forecast errors, work reasonably well compared to more sophisticated combination schemes.

The vast majority of studies on combining multiple forecasts have dealt with point forecasting, even though point forecasts (without associated measures of uncertainty) provide insufficient information for decision-making. The simple arithmetic average of forecasts based on equal weights stands out as the most popular and surprisingly robust combination rule \citep[see][]{Bunn1985-vo,Clemen1986-pd,Stock2003-sp,Genre2013-ut}, and can be effortlessly implemented.

An early example of an equally weighted combination is from the M-competition, the first forecasting competition run by Spyros Makridakis and Mich{\`e}le Hibon, involving 1001 time series; see \citet{Makridakis1982-hb} and \citet{Hyndman2020-bh} for more details of the competition. \citet{Makridakis1982-hb} reported that the simple average outperformed the individual forecasting models. \citet{Clemen1989-fb} provided an extensive bibliographical review of the early work on the combination of forecasts, and then addressed the issue that the arithmetic means often dominate more refined forecast combinations. \citet{Makridakis1983-hg} concluded empirically that a larger number of individual methods included in the simple average scheme would help improve the accuracy of combined forecasts and reduce the variability associated with the selection of methods. \citet{Palm1992-im} concisely summarized the advantages of adopting simple averaging into three aspects: (i) combination weights are equal and do not have to be estimated; (ii) simple averaging significantly reduces variance and bias by averaging out individual bias in many cases; and (iii) simple averaging should be considered when the uncertainty of weight estimation is taken into account. Additionally, \citet{Timmermann2006-en} pointed out that the outstanding average performance of simple averaging depends strongly on model instability and the ratio of forecast error variances associated with different forecasting models.

More attention has been given to other strategies, including using the median and mode, as well as trimmed and winsorized means \citep[e.g.,][]{Chan1999-io,Stock2004-rq,Genre2013-ut,Jose2014-uh,Grushka-Cockayne2017-dj}, due to their robustness in the sense of being less sensitive to extreme forecasts than a simple average \citep{Lichtendahl2020-ut}. For example, the early work of \citet{Galton1907-vox} observed that the ``middlemost'' of 787 estimates of an ox's weight is within nine pounds of the ox's actual weight, and thus advocated for the median forecast as the ``vox populi'' \citep{Galton1907-one}. However, there is little consensus in the literature on whether the mean or the median of individual forecasts performs better in terms of point forecasting \citep{Kolassa2011-ai}. Specifically, \citet{McNees1992-qc} found no significant difference between the mean and the median, while the results of \citet{Stock2004-rq} supported the mean and \citet{Agnew1985-dj,Galton1907-vox} recommended the median. \citet{Jose2008-vm} studied the forecasting performance of the mean and median, as well as the trimmed and winsorized means. Their results suggested that the trimmed and winsorized means are appealing, particularly when there is a high level of variability among the individual forecasts, because of their simplicity and robust performance. \citet{Kourentzes2014-hs} compared empirically the mean, mode and median combination operators based on kernel density estimation, and found that the three operators deal with outlying extreme values differently, with the mean being the most sensitive and the mode operator the least. Based on these experimental results, they recommended further investigation of the use of the mode and median operators, which have been largely overlooked in the relevant literature.

Compared to various complicated combination approaches and machine learning algorithms, simple combinations seem outdated and uncompetitive in the big data era. However, the results from the recent M4 competition \citep{Makridakis2020-hu} showed that simple combinations continue to achieve relatively good forecasting performance and are still competitive. Specifically, a simple equal-weight combination ranked the third for yearly time series \citep{Shaub2019-on} and a median combination of four simple forecasting models achieved the sixth place for point forecasting \citep{Petropoulos2020-fp}. \citet{Genre2013-ut} encompassed a variety of combination methods in the case of forecasting GDP growth and the unemployment rate. They found that the simple average sets a tough benchmark, with few combination schemes outperforming it. Moreover, simple combinations have a lower computational burden and can be implemented more efficiently than alternatives. Therefore, simple combination rules have been consistently the choice of many researchers and practitioners, and provide a challenging benchmark to measure the effectiveness of the newly proposed weighted forecast combination algorithms \citep[e.g.,][]{Makridakis2000-he,Stock2004-rq,Makridakis2020-hu,Montero-Manso2020-tq,Kang2020-rl,Wang2021-un}.

Despite the ease of implementing simple combination schemes, their success still depends largely on the choice of the forecasts to be combined. Intuitively, we prefer that the component forecasts fall on opposite sides of the truth (the realization) \citep{Bates1969-yj,Larrick2006-sr}, so that the forecast errors tend to cancel each other out. However, this rarely occurs in practice, as the component forecasts are usually trained based on overlapping information sets and use similar forecasting methods. If all component forecasts are established similarly based on the same, or highly overlapping sets of information, forecast combinations are unlikely to be beneficial for the improvement of forecast accuracy. \citet{Mannes2014-dl} and \citet{Lichtendahl2020-ut} emphasized two critical issues concerning the performance of simple combination rules: one for the level of accuracy (or expertise) of the forecasts in the pool and another for diversity among individual forecasts. Involving forecasts with low accuracy in the pool can decrease the combination performance. Additionally, a high degree of diversity among component models facilitates the achievement of the best possible forecast accuracy from simple combinations \citep{Thomson2019-al}. In conclusion, simple, easy-to-use combination rules can provide good and robust forecasting performance, especially when properly considering issues such as the accuracy and diversity of the individual forecasts to be combined.

\subsection{Linear combinations}
\label{sec:linear_comb}

Despite the simplicity and performance of simple combination rules, it makes sense to assign greater weight to the most accurate forecast methods. But how to choose those weights? The problem of point forecast combinations can be defined as seeking a one-dimensional aggregator that integrates an $N$-dimensional vector of $h$-step-ahead forecasts involving the information up to time $T$, $\hat{\bm{y}}_{T+h|T}=\left(\hat{y}_{T+h|T, 1}, \hat{y}_{T+h|T, 2}, \dots, \hat{y}_{T+h|T, N}\right)^{\prime}$, into a single combined $h$-step-ahead forecast $\tilde{y}_{T+h|T}=C\left(\hat{\bm{y}}_{T+h|T} ; \bm{w}_{T+h|T}\right)$, where $N$ is the number of forecasts to be combined and $\bm{w}_{T+h|T}$ is an $N$-dimensional vector of combining weights. The class of combination methods represented by the mapping, $C$, comprises linear and nonlinear combinations, as well as series-specific and cross-learning combinations. Additionally, the combination weights can be static or time-varying along the forecasting horizon. Below we discuss in detail the use of various approaches to determining combination weights associated with individual forecasts.

Typically, the combined forecast is constructed as a linear combination of the individual forecasts, which can be written as
\begin{align*}
  \tilde{y}_{T+h|T}=\bm{w}_{T+h|T}^{\prime} \hat{\bm{y}}_{T+h|T},
\end{align*}
where $\bm{w}_{T+h|T}=\left(w_{T+h|T, 1}, \dots, w_{T+h|T, N}\right)^{\prime}$ is an $N$-dimensional vector of linear combination weights assigned to $N$ individual forecasts.

\subsubsection*{Optimal weights}

The seminal work of \citet{Bates1969-yj} proposed a method to find the so-called ``optimal'' weights by minimizing the variance of the combined forecast error, and discussed only combinations of pairs of forecasts. \citet{Newbold1974-lp} then extended the method to combinations of more than two forecasts. Specifically, if the individual forecasts are unbiased and their error variances are consistent over time, then the combined forecast obtained by a linear combination will also be unbiased. Differentiating with respect to $\bm{w}_{T+h|T}$ and solving the first order condition, the variance of the combined forecast error is minimized by taking
\begin{align}
  \label{eq:weight_opt}
  \bm{w}_{T+h|T}^{\text{opt}}=\frac{\bm{\Sigma}_{T+h|T}^{-1}\bm{1}}{\bm{1}^{\prime} \bm{\Sigma}_{T+h|T}^{-1} \bm{1}},
\end{align}
where $\bm{\Sigma}_{T+h|T}$ is the $N \times N$ covariance matrix of the $h$-step forecast errors and $\bm{1}$ is an $N$-dimensional unit vector. This is implemented, for example, in the \proglang{R} package \pkg{ForecastComb} \citep{rForecastComb}. In practice, the elements of the covariance matrix $\bm{\Sigma}_{T+h|T}$ are usually unknown and need to be estimated.

It follows that if $\bm{w}_{T+h|T}$ is determined by Equation~\eqref{eq:weight_opt}, one can identify a combined forecast $\tilde{y}_{T+h|T}$ with no greater error variance than the minimum error variance of all individual forecasts. The fact was further explored by \citet{Timmermann2006-en} to illustrate the diversification gains offered by forecast combinations, by simply considering combinations of pairs of forecasts. Under mean squared error (MSE) loss, \citet{Timmermann2006-en} characterized the general solution of the optimal linear combination weights by assuming a joint Gaussian distribution of the outcome $y_{T+h}$ and available forecasts $\hat{\bm{y}}_{T+h|T}$.

The loss assumed in \citet{Bates1969-yj} and \citet{Newbold1974-lp} is quadratic and symmetric. \citet{Elliott2004-dz} examined forecast combinations under more general loss functions accounting for asymmetries as well as skewed forecast error distributions. They demonstrated that the optimal combination weights strongly depend on the degree of asymmetry in the loss function and skewness in the underlying forecast error distributions. Subsequently, \citet{Patton2007-zo} demonstrated that the properties of optimal forecasts established under MSE loss are not generally robust under more general assumptions about the loss function. In addition, the properties of optimal forecasts were generalized to consider asymmetric loss and nonlinear data generating processes.

\subsubsection*{Regression-based weights}

The seminal work by \citet{Granger1984-jc} provided an important impetus for approximating the ``optimal'' weights under a linear regression framework. They recommended the strategy that the combination weights can be estimated by ordinary least squares (OLS) in regression models having the vector of past observations as the response variable and the matrix of past individual forecasts as the predictor variables. Three alternative approaches imposing various possible restrictions were considered
\begin{align}
   & y_{T+h}=\bm{w}_{T+h|T}^{\prime} \hat{\bm{y}}_{T+h|T}+\varepsilon_{T+h}, \quad s.t. \quad \bm{w}^{\prime}\bm{1}=1, \label{eq:weight_gr1} \\
   & y_{T+h}=\bm{w}^{\prime}_{T+h|T} \hat{\bm{y}}_{T+h|T}+\varepsilon_{T+h}, \\
   & y_{T+h}=w_{T+h|T, 0}+\bm{w}_{T+h|T}^{\prime} \hat{\bm{y}}_{T+h|T}+\varepsilon_{T+h}. \label{eq:weight_gr3}
\end{align}
The \proglang{R} package \pkg{ForecastComb} \citep{rForecastComb} provides the corresponding implementations. The constrained OLS estimation of the regression in Equation~\eqref{eq:weight_gr1}, in which the constant is omitted and the weights are constrained to sum to one, yields results identical to the ``optimal'' weights proposed by \citet{Bates1969-yj}. \citet{Granger1984-jc} further suggested that the unrestricted OLS regression in Equation~\eqref{eq:weight_gr3}, which allows for a constant term and does not require the weights to sum to one, is superior to the popular ``optimal'' method regardless of whether the constituent forecasts are biased. However, \citet{De_Menezes2000-vd} argue that when using the unrestricted regression, one needs to consider the stationarity of the series being forecast, the possible presence of serial correlation in forecast errors \citep[see also][]{Diebold1988-sx,Edward_Coulson1993-db}, and the issue of multicollinearity.

Generalizations of the combination regressions have been considered in a large body of literature. \citet{Diebold1988-sx} exploited serial correlated errors in the least squares framework by characterizing the combined forecast errors as autoregressive moving average (ARMA) processes, leading to improved combined forecasts. \citet{Gunter1992-go} and \citet{Aksu1992-lb} provided an empirical analysis to compare the performance of various combination strategies, including the simple average, the unrestricted OLS regression, the restricted OLS regression where the weights are constrained to sum to unity, and the restricted OLS regression where the weights are constrained to be nonnegative. The results revealed that constraining weights to be nonnegative is at least as robust and accurate as the simple average and yields superior results compared to other combinations based on a regression framework. \citet{Conflitti2015-fq} addressed the problem of determining the combination weights by imposing both restrictions (that the weights should be nonnegative and sum to one), which turns out to be a special case of a LASSO regression. \citet{Edward_Coulson1993-db} found that allowing a lagged dependent variable in forecast combination regressions can achieve improved performance. Instead of using the quadratic loss function, \citet{Nowotarski2014-ev} applied the absolute loss function in the unrestricted regression, also implemented in the \pkg{ForecastComb} package for \proglang{R}, to yield the least absolute deviation regression which is more robust to outliers than OLS combinations.

Forecast combinations using changing weights have also been developed to solve various types of structural changes in constituent forecasts. For instance, \citet{Diebold1987-go} explored rolling weighted least squares as well as time-varying parameter techniques in the basic regression framework, including both deterministic and stochastic time-varying parameters. Specifically, the combination weights are either described as deterministic nonlinear (polynomial) functions of time or allowed to involve random variation. They showed, via numerical examples based on various types of structural change in the constituent forecasts, that time-varying weights substantially help in improving forecasting ability in the presence of instabilities. \citet{Deutsch1994-ob} allowed the combination weights to evolve immediately or smoothly using switching regression models and smooth transition regression models. \citet{Terui2002-df} generalized the regression method by incorporating time-varying coefficients that are assumed to follow a random walk process. The generalized model can be interpreted as a state space model and then estimated using Kalman filter updating. Following the spirit of \citet{Terui2002-df}, \citet{Raftery2010-qe} achieved an accelerated inference process by using forgetting factors in the recursive Kalman filter updating.

Researchers have also worked on including many forecasts in a regression framework to take advantage of many models. However, \citet{Chan1999-io} examined a wide range of combination methods and showed that OLS combinations have very poor performance when $N$ (the number of forecasts to be combined) is very large. Factor methods are a common way of condensing information when modeling and forecasting. They have also been used explicitly in forecast combination settings, and are especially attractive when the number of forecasts to be combined is very large ($N > T$); see, e.g., \citet{Chan1999-io} for a dynamic factor model framework for forecast combinations. The common factors in approximate dynamic factor models can be estimated by principal components \citep{Stock1999-fi}. Principal components regression (PCR) is typically motivated as an ad hoc tool for the solution of multicollinearity. \citet{Chan1999-io} and \citet{Stock2004-rq} explicitly applied PCR to forecast combinations, resulting in a two-step procedure. The first step extracts the principal components, while the second step produces the final forecasts utilizing OLS regression. The superiority of PCR over OLS combinations was also supported by \citet{Rapach2008-jh} and \citet{Poncela2011-vz}. In turn, these methods relate to the question of whether one should forecast with variables (competing point forecasts in our paper's context), factors (extracted from the $N$ competing forecasts), or both; see, e.g., \citet{Castle2013-fv} for a detailed discussion.

In large $N$ cases, given the estimation problems that arise when $N > T$, researchers also frequently relate forecast combinations to shrinkage-type approaches (whether frequentist or Bayesian) that facilitate estimation of the forecast combination regression even when $N > T$; e.g. see \citet{Stock2004-rq}. \citet{Diebold2019-ml} considered methods for selection and shrinkage in regression-based forecast combinations to address the estimation problem. They shed light on how machine learning can be used to optimally combine a large set of forecasts by introducing a LASSO-based procedure that consists of two steps. The first step involves setting some combination weights to zero using LASSO, and the second step shrinks the combination weights of the survivors toward equal weights. Additionally, \citet{Aiolfi2006-rh} argued in favor of clustering the individual forecasts using the $k$-means clustering algorithm based on their historical performance. For each cluster, a pooled (average) forecast is computed, which precedes the estimation of combination weights for the constructed clusters.

\subsubsection*{Performance-based weights}

Estimation errors in the ``optimal'' weights and regression-based weights tend to be particularly large due to difficulties in properly estimating the covariance matrix $\bm{\Sigma}_{T+h|T}$, especially in situations with many forecasts to combine. Instead, \citet{Bates1969-yj} suggested weighing the constituent forecasts in inverse proportion to their historical performance, ignoring mutual dependence. In follow-up studies, \citet{Newbold1974-lp} and \citet{Winkler1983-ra} generalized this idea in the sense of considering more time series, more individual forecasts, and multiple forecast horizons. Their extensive results demonstrated that combinations ignoring correlations are more successful than those attempting to take account of correlations, and consequently reconfirmed \citeapos{Bates1969-yj} argument that correlations can be poorly estimated in practice and should be ignored when calculating combination weights.

Let $\bm{e}_{T+h|T}=\bm{1} y_{T+h}-\hat{\bm{y}}_{T+h|T}$ be the $N$-dimensional vector of $h$-step forecast errors computed from the individual forecasts. Then the five procedures suggested in \citet{Bates1969-yj} for estimating the combination weights when $\bm{\Sigma}_{T+h|T}$ is unknown, are extended to the general case as follows:
\begin{align}
  & w_{T+h|T, i}^{\text{bg1}}=\frac{\left( \sum_{t=T-\nu+1}^{T} e_{t|t-h, i}^{2} \right)^{-1}}{\sum_{j=1}^{N}\left(\sum_{t=T-\nu+1}^{T} e_{t|t-h, j}^{2}\right)^{-1}}; \label{eq:weight_bg1} \\
  & \bm{w}_{T+h|T}^{\text{bg2}}=\frac{\hat{\bm{\Sigma}}_{T+h|T}^{-1}\bm{1}}{\bm{1}^{\prime} \hat{\bm{\Sigma}}_{T+h|T}^{-1} \bm{1}}, \quad \text{where} \quad (\hat{\bm{\Sigma}}_{T+h|T})_{i, j}=\nu^{-1} \sum_{t=T-\nu+1}^{T} e_{t|t-h, i} e_{t|t-h, j};   \\
  & w_{T+h|T, i}^{\text{bg3}}=\alpha \hat{w}_{T+h-1|T-1, i} + (1-\alpha) \frac{\left( \sum_{t=T-\nu+1}^{T} e_{t|t-h, i}^{2} \right)^{-1}}{\sum_{j=1}^{N}\left(\sum_{t=T-\nu+1}^{T} e_{t|t-h, j}^{2}\right)^{-1}}, \quad \text{where} \quad 0<\alpha<1;   \\
  & w_{T+h|T, i}^{\text{bg4}}=\frac{\left( \sum_{t=1}^{T} \gamma^{t} e_{t|t-h, i}^{2} \right)^{-1}}{\sum_{j=1}^{N}\left(\sum_{t=1}^{T} \gamma^{t} e_{t|t-h, j}^{2}\right)^{-1}}, \quad \text{where} \quad \gamma \geq 1; \label{eq:weight_bg4}  \\
  & \bm{w}_{T+h|T}^{\text{bg5}}=\frac{\hat{\bm{\Sigma}}_{T+h|T}^{-1}\bm{1}}{\bm{1}^{\prime} \hat{\bm{\Sigma}}_{T+h|T}^{-1} \bm{1}}, \quad \text{where} \quad (\hat{\bm{\Sigma}}_{T+h|T})_{i, j}=\frac{\sum_{t=1}^{T} \gamma^{t} e_{t|t-h, i} e_{t|t-h, j}}{\sum_{t=1}^{T} \gamma^{t}} \quad \text{and} \quad \gamma \geq 1.
\end{align}
These weighting schemes differ in the factors, as well as the choice of the parameters, $\nu$, $\alpha$, and $\gamma$. Correlations across forecast errors are either ignored by treating the covariance matrix $\bm{\Sigma}_{T+h|T}$ as a diagonal matrix or estimated via the usual sample estimator (which may lead to quite unstable estimates of $\bm{\Sigma}_{T+h|T}$ given highly correlated forecast errors). Some estimation schemes suggest computing or updating the relative performance of individual forecasts over rolling windows of the most recent $\nu$ observations, while others base the weights on exponential discounting with higher values of $\gamma$ giving larger weights to more recent observations. Consequently, these weighting schemes are well adapted to allow a non-stationary relationship between the individual forecasting procedures over time \citep{Newbold1974-lp}. However, they tend to increase the variance of the parameter estimates and work quite poorly if the data generating process is truly covariance stationary \citep{Timmermann2006-en}.

A broader set of combination weights based on the relative performance of individual forecasting techniques has been developed and examined in a series of studies. For example, \citet{Stock1998-np} generalized the rolling window scheme in Equation~\eqref{eq:weight_bg1} in the sense that the weights on the individual forecasts are inversely proportional to the $k$th power of their MSE. The weights with $k = 0$ correspond to assigning equal weights to all forecasts, while more weights are placed on the best performing forecasts by considering $k \geq 1$. Other forms of forecast error measures, such as the root mean squared error (RMSE) and the symmetric mean absolute percentage error (sMAPE), have also been considered to lead to performance-based combination weights \citep[e.g.,][]{Nowotarski2014-ev,Pawlikowski2020-hm}. A weighting scheme with the weights depending inversely on the exponentially discounted errors was proposed by \citet{Stock2004-rq} as an upgraded version of the scheme in Equation~\eqref{eq:weight_bg4}, and was used in several subsequent studies \citep[e.g.,][]{Clark2010-jx,Genre2013-ut} to achieve gains from combining forecasts. The pseudo out-of-sample performance used in these weighting schemes is commonly computed based on rolling or recursive (expanding) windows \citep[e.g.,][]{Stock1998-np,Clark2010-jx,Genre2013-ut}. It is natural to adopt rolling windows in estimating the weights to deal with structural changes, but the window length should not be too short without the estimates of the weights becoming too noisy \citep{Baumeister2015-ft}.

Compared to constructing the weights directly using historical forecast errors, a new form of combinations that is more robust and less sensitive to outliers was introduced based on the ``ranking'' of individual forecasts. Again this kind of combination ignores correlations among forecast errors. The simplest and most commonly used method in the class is to use the median forecast as the output. \citet{Aiolfi2006-rh} constructed the weights proportional to the inverse of performance ranks (sorted according to increasing order of forecast errors), which were later employed by \citet{Andrawis2011-kb} for tourism demand forecasting. The \proglang{R} package \pkg{ForecastComb} \citep{rForecastComb} provides tools for rank-based combinations. Another weighting scheme which attaches a weight proportional to $\exp(\beta(N+1-i))$ to the $i$th ordered constituent forecast was adopted in \citet{Yao2008-or} and \citet{Donate2013-lq} to combine forecasts obtained from artificial neural networks (ANNs), where $\beta$ is a scaling factor. However, as mentioned by \citet{Andrawis2011-kb}, this class of combination methods limits the weights to only a discrete set of possible values.

\subsubsection*{Criteria-based weights}

Information criteria, such as Akaike's information criterion \citep[AIC,][]{Akaike1974-ya}, the corrected Akaike information criterion \citep[AICc,][]{Sugiura1978-xm}, and the Bayesian information criterion \citep[BIC,][]{Schwarz1978-cz}, are often used for model selection in forecasting. However, choosing a single model out of the candidate model pool may be misleading because of the information loss from the alternative models. An alternative approach proposed by \citet{Burnham2002-us} is to combine multiple models based on information criteria to mitigate the risk of selecting a single model. It is also worth mentioning that the \proglang{R} packages \pkg{MuMIn} \citep{rMuMIn} and \pkg{mmSAR} \citep{rmmSAR} have been developed to perform model selection and multimodel averaging based on the use of information-theoretic approaches introduced by \citet{Burnham2002-us}.

One such common approach is using Akaike weights. Specifically, in light of the fact that AIC estimates the Kullback-Leibler distance \citep{Kullback1951-hl} between a model and the true data generating process, differences in the AIC can be used to weight different models, providing a measure of the evidence for supporting a given model relative to other constituent models. Given $N$ individual models, the Akaike weight of model $i$ can be derived as:
\begin{align*}
    & w_{T+h|T, i}^{\text{aic}}=\frac{\exp (-0.5 \Delta \mathrm{AIC}_{i})}{\sum_{k=1}^{N} \exp \left(-0.5 \Delta \mathrm{AIC}_{k}\right)}, \\
  \text{where}\qquad
    & \Delta \mathrm{AIC}_{i}=\mathrm{AIC}_{i}-\min _{k \in \{1,2,\cdots,N\}} \mathrm{AIC}(k).
\end{align*}
Akaike weights calculated in this manner can be interpreted as the probability that a given model performs best at approximating the unknown data generating process, given the model set and the available and historical data \citep{Kolassa2011-ai}. Similar weights from AICc, BIC, and other variants with different penalties, can be derived analogously.

The outstanding performance of weighted combinations based on information criteria has been supported in several studies. For instance, \citet{Kolassa2011-ai} used weights derived from AIC, AICc and BIC to combine exponential smoothing forecasts, and obtained superior accuracy over selecting a model using the same information criteria. A similar strategy was adopted by \citet{Petropoulos2018-fw} to separately explore the benefits of bootstrap aggregation (bagging) for time series forecasting. Additionally, an empirical study by \citet{Petropoulos2018-ad} showed that a weighted combination based on AIC improves the performance of the statistical benchmark they used.

\subsubsection*{Bayesian weights}

Some effort has been directed towards Bayesian approaches to updating forecast combination weights in the face of new information gleaned from various sources. Recall that obtaining reliable estimates of the covariance matrix $\bm{\Sigma}$ (the time and horizon subscripts are dropped for simplicity) of forecast errors is a major challenge in practice regardless of whether correlations among forecast errors are ignored or not. With this in mind, \citet{Bunn1975-vz} suggested the idea of Bayesian combinations on the basis of the probability of each forecasting model performing the best on any given occasion. Considering the beta and the Dirichlet distributions as the conjugate priors for the binomial and multinomial processes respectively, the suggested non-parametric method performs well when there is relatively little past data by means of attaching prior subjective probabilities to individual forecasts \citep{Bunn1985-vo,De_Menezes2000-vd}. \citet{Oller1978-wx} presented another approach to using subjective probability in a Bayesian updating scheme based on the self-scoring weights proportional to the evaluation of the expert's forecasting ability.

A different strand of research has also advocated the incorporation of prior information into the estimation of combination weights, but with the weights being shrunk toward some prior mean under a regression-based combination framework \citep{Newbold2002-wa}. Assuming that the vector of forecast errors is normally distributed, \citet{Clemen1986-pd} developed a Bayesian approach with the conjugate prior for $\bm{\Sigma}$ represented by an inverted Wishart distribution with covariance matrix $\bm{\Sigma}_{0}$ and scalar degrees of freedom $\nu_{0}$. Again we drop time and horizon subscripts for simplicity. If the last $T$ observations are used to estimate $\bm{\Sigma}$, the combination weights derived from the posterior distribution for $\bm{\Sigma}$ are
\begin{align*}
  \bm{w}^{\text{cw}}=\frac{\bm{\Sigma}^{*}\bm{1}}{\bm{1}^{\prime} \bm{\Sigma}^{*} \bm{1}},
\end{align*}
where $\bm{\Sigma}^{*}=\big(\nu_{0} \bm{\Sigma}_{0}^{-1}+T \hat{\bm{\Sigma}}^{-1}\big) /(\nu_{0}+T)$ is the precision matrix and $\hat{\bm{\Sigma}}$ is the sample covariance matrix.
Compared to estimating $\bm{\Sigma}$ using the standard sample covariance estimator or treating it as a diagonal matrix, the proposed approach provides a more stable estimation and allows for correlations between methods. The subsequent work by \citet{Diebold1990-fk} allowed the incorporation of the standard normal-gamma conjugate prior by considering a normal regression-based combination
\begin{align*}
  \bm{y}=\hat{\bm{Y}} \bm{w}+\bm{\varepsilon}, \quad \bm{\varepsilon} \sim N\left(\bm{0}, \bm{\sigma}^{2} \bm{I}\right),
\end{align*}
where $\bm{y}$ and $\bm{\varepsilon}$ are $T$-dimensional vectors of historical data and residuals, respectively, and $\hat{\bm{Y}}$ is the $T \times N$ matrix of constituent one-step forecasts. This approach results in estimated combination weights that can be viewed as a matrix-weighted average of those for the two polar cases (least squares and prior weights), and it can provide a rational transition between the subjective and data-based estimation of the combination weights. In light of the fact that Bayesian approaches have been mostly employed to construct combinations of probability forecasts, we will elaborate on other newly developed methods of determining combination weights from a foundational Bayesian perspective in Section~\ref{sec:bayesian_comb}.

\subsection{Nonlinear combinations}
\label{sec:nonlinear_comb}

Linear combination approaches implicitly assume a linear dependence between constituent forecasts and the variable of interest \citep{Donaldson1996-um,Freitas2006-fn}, and may not result in the best forecast \citep{Ming_Shi1999-vs}, especially if the individual forecasts come from nonlinear models or if the true relationship between combination members and the best forecast is characterized by nonlinear systems \citep{Babikir2016-xz}. In such cases, it is natural to relax the linearity assumption and consider nonlinear combination schemes of higher complexity; these have received very limited research attention so far.

As \citet{Timmermann2006-en} identified, two types of nonlinearities can be incorporated in forecast combinations. One involves nonlinear functions of the individual forecasts, but with the unknown parameters of the combination weights given in the linear form. The other allows a more general combination with nonlinearities directly considered in the combination parameters. Neural networks are often employed to estimate the nonlinear mapping because they offer the potential of learning the underlying nonlinear relationship between the future outcome and individual forecasts. The design of a neural network model is nevertheless time-consuming, and sometimes leads to overfitting and poor forecasting performance as more parameters need to be estimated.

\citet{Donaldson1996-um} used ANNs to obtain the combined forecasts $\tilde{y}_{T+h|T}$ by the following form
\begin{align}
  \tilde{y}_{T+h|T} & =
    \beta_{0}+\sum_{j=1}^{k} \beta_{j} \hat{y}_{T+h|T, j}+\sum_{i=1}^{p} \delta_{i} g(\bm{z}_{T+h|T} \bm{\gamma}_{i}), \label{eq:nonlinear_sl} \\
  g(\bm{z}_{T+h|T} \bm{\gamma}_{i}) & =
    \left(1+\exp \bigg\{-\Big(\gamma_{0, i}+\sum_{j=1}^{N} \gamma_{1, j} z_{T+h|T, j}\Big)\bigg\}\right)^{-1},
\end{align}
where $z_{T+h|T, j} = (\hat{y}_{T+h|T, j}-\bar{y}) / \hat{\sigma}$, $\bar{y}$ and $\hat{\sigma}$ denote the in-sample mean and in-sample standard deviation respectively, $k \in \{0,N\}$, and $p \in \{0,1,2,3\}$. This approach permits special cases of both purely linear combinations ($k=N$ and $p=0$) and nonlinear combinations ($k=0$ and $p\neq 0$). Building on this, \citet{Harrald1997-gd} proposed to evolve ANNs and demonstrated their utility, but only using a single time series. \citet{Krasnopolsky2012-xu} and \citet{Babikir2016-xz} employed neural network approaches with various activation functions to approximate the nonlinear dependence of individual forecasts and achieve nonlinear mapping, resulting in variants of Equation~\eqref{eq:nonlinear_sl}. The empirical results of nonlinear combinations from these studies generally dominate those from traditional linear combination strategies, such as simple average, OLS weights, and performance-based weights. However, the empirical evidence provided is based on fewer than ten time series, possibly hand-picked to lead to this result. Additionally, these nonlinear combination methods suffer from other drawbacks including the neglect of correlations among forecast errors, the instability of parameter estimation, and the multicollinearity caused by the overlap in the information sets used to produce individual forecasts. Thus, the performance of nonlinear combinations relative to linear combinations needs further investigation.

Some researchers have sought to construct nonlinear combinations via the inclusion of an additional nonlinear term to cope with the case where the individual forecast errors are correlated. The combination mechanism can be generalized to the following form
\begin{align*}
  \tilde{y}_{T+h|T} =\beta_{0}+\sum_{j=1}^{N} \beta_{j} \hat{y}_{T+h|T, j}+\sum_{i,j=1 \atop i<j}^{N}\pi_{ij}v_{i j},
\end{align*}
where $v_{ij}$ is some nonlinear combination of forecasts $i$ and $j$. In this way, the general framework for linear combinations is extended to deal with nonlinearities.

For example, \citet{Freitas2006-fn} defined $v_{i j}$ as the product of individual forecasts from different models, $v_{ij} = \hat{y}_{T+h|T, i} \hat{y}_{T+h|T, j}$, while \citet{Adhikari2012-ur} took into account the linear correlations among the forecast pairs by including the term, $v_{ij} = (\hat{y}_{T+h|T, i}-\bar{y}_{i})(\hat{y}_{T+h|T, j}-\bar{y}_{j})/(\sigma_{i}\sigma_{j})^2$, where $\bar{y}_{i}$ and $\sigma_{i}$ are the mean and standard deviation of the $i$th model, respectively. Moreover, \citet{Adhikari2015-bb} defined the nonlinear term using $v_{ij} = \left(\hat{z}_{i}-m_{i j} \hat{z}_{j}\right)\left(\hat{z}_{j}-m_{j i} \hat{z}_{i}\right)$, where $\hat{z}_{i}$ denotes the standardized $i$th individual forecast using the mean $\bar{y}_{i}$ and standard deviation $\sigma_{i}$, and the term $m_{i j}$ denotes the degree of mutual dependency between the $i$th and $j$th individual forecasts.

Clearly, combining forecasts nonlinearly requires further research. In particular, the forecasting performance of the various proposed nonlinear combination schemes should be properly investigated with a large, diverse collection of time series datasets along with appropriate statistical inference. There is also a need to develop nonlinear combination approaches that take account of correlations across forecast errors and the multicollinearity of forecasts.

\subsection{Combining by learning}
\label{sec:comb_learn}

Stacked generalization \citep[stacking,][]{Wolpert1992-if} provides a strategy to adaptively combine the available forecasting models. Stacking is frequently employed on a wide variety of classification tasks \citep{Zhou2012-cy}; in the time series forecast context, it uses the concept of meta-learning to boost forecasting accuracy beyond that achieved by any of the individual models. Stacking is a general framework that comprises at least two levels. The first level involves training the individual forecasting models using the original data, while the second and any subsequent levels utilize an additional ``meta-model'', using the prior level forecasts as inputs to form a set of forecasts. Thus, the stacking approach to forecast combinations weights individual forecasts adaptively using meta-learning processes.

There are many ways to implement the stacking strategy. Its primary implementation is to combine individual models in a series-by-series fashion. Individual forecasting models in the method pool are trained using only data of the single series they are going to forecast, while their forecast outputs are subsequently fed to a meta-model tailored for the target series to calculate the combined forecasts. This means that $n$ meta-models are required for $n$ separate time series data. Unsurprisingly, regression-based weight combinations discussed in Section~\ref{sec:linear_comb} \citep[e.g.,][]{Granger1984-jc,Gunter1992-go} fall into this category and can be viewed as the most simple, common learning algorithm used in stacking. Instead of applying multiple linear regressions, \citet{Moon2020-ls} suggested a PCR model as the meta-model predominantly due to its desirable characteristics such as dimensionality reduction and avoidance of multicollinearity between the input forecasts of individual models. Similarly, LASSO regression, ANN, wavelet neural network (WNN), and support vector regression (SVR) can be conducted in a series-by-series fashion to achieve the same goal \citep[e.g.,][]{Donaldson1996-um,Conflitti2015-fq,Ribeiro2019-wk,Ribeiro2020-mj}. One could use an expanding or rolling window method to ensure that enough individual forecasts are generated for the training of meta-models. Time series cross-validation, also known as ``evaluation on a rolling forecasting origin'' \citep{Hyndman2021-tx}, is also recommended in the training procedures for both individual models and meta-models to help with the parameter estimation. Nevertheless, stacking approaches implemented in a series-by-series fashion still suffer from some limitations such as requiring a long computation time and long time series, and inefficiently using the training data.

An alternative way to perform the stacking strategy sheds some light on the potential of cross-learning. Specifically, the meta-model is trained using information derived from multiple time series rather than employing only a single series, thus various patterns can be captured along different series. The M4 competition organized by Spyros \citet{Makridakis2020-hu}, comprising $100,000$ time series, recognized the benefits of cross-learning in the sense that the top three performing methods of the competition utilize the information across the whole dataset rather than a single series. Cross-learning can therefore be identified as a promising strategy to boost forecasting accuracy, at least when appropriate strategies for extracting information from large, diverse time series datasets are adopted \citep{Kang2020-sa,Semenoglou2020-xx}. \citet{Zhao2020-ep} trained a neural network model across the M4 competition dataset to learn how to combine individual models in the method pool. They adopted the temporal holdout strategy to generate the training dataset and utilized only the out-of-sample forecasts produced by standard individual models as the input for the neural network model.

An increasing stream of studies has shown that time series features characterizing each series in a dataset, provide valuable information for forecast combinations in a cross-learning fashion, leading to an extension of stacking. Numerous software packages have been developed for time series feature extraction, including the \proglang{R} packages \pkg{feasts} \citep{rfeasts} and \pkg{tsfeatures} \citep{rtsfeatures}, the \proglang{Python} packages \pkg{Kats} \citep{pKats}, \pkg{tsfresh} \citep{Christ2018-vi} and \pkg{TSFEL} \citep{Barandas2020-vr}, the \proglang{Matlab} package \pkg{hctsa} \citep{Fulcher2017-uf}, and the \proglang{C}-coded package \pkg{catch22} \citep{Lubba2019-ds}. These sets of time series features were empirically evaluated by \citet{Henderson2021-gl}.

The pioneering work by \citet{Collopy1992-ey} developed a rule base consisting of $99$ rules to combine forecasts from four statistical models using $18$ time series features. \citet{Petropoulos2014-uy} identified the main determinants of forecasting accuracy through an empirical study involving $14$ forecasting models and seven time series features. The findings can provide useful information for forecast combinations. More recently, \citet{Montero-Manso2020-tq} introduced a Feature-based FORecast Model Averaging (FFORMA) approach available in the \proglang{R} package \pkg{M4metalearning} \citep{rfforma}, which employs $42$ statistical features (implemented using the \proglang{R} package \pkg{tsfeatures}) to estimate the optimal weights for combining nine different traditional models trained per series based on an XGBoost model. The FFORMA method reported the second-best forecasting accuracy in the M4 competition. Additionally, \citet{Ma2021-np} highlighted the potential of convolutional neural networks as a meta-model to link the learnt features with a set of combination weights. \citet{Li2020-od} extracted time series features automatically with the idea of time series imaging, then these features were used for forecast combinations. \citet{Gastinger2021-ey} demonstrated the value of a collection of combination methods on a large and diverse amount of time series from the M3 \citep{Makridakis2000-he}, M4, M5 \citep{Makridakis2020-fn} datasets and FRED datasets\footnote{The FRED (Federal Reserve Economic Data) dataset is openly available at \url{https://fred.stlouisfed.org}.}. In light of the fact that it is not clear which combination strategy should be selected, they introduced a meta-learning step to select a promising subset of combination methods for a newly given dataset based on extracted features.

In addition to the time series features extracted from the historical data, it is crucial to look at the diversity of the individual model pool in the context of forecast combinations \citep{Batchelor1995-ps,Thomson2019-al,Atiya2020-ge,Lichtendahl2020-ut}. An increase in diversity among forecasting models has the potential to improve the accuracy of their combination. In this respect, features measuring the diversity of the method pool should be included in the feature pool to provide additional information possibly relevant to combining models. \citet{Lemke2010-wn} calculated six diversity features and created an extensive feature pool describing both the time series and the individual method pool. Three meta-learning algorithms were implemented to link knowledge on the performance of individual models with the extracted features, and to improve forecasting performance. \citet{Kang2021-ol} utilized a group of features only measuring the diversity across the candidate forecasts to construct a forecast combination model mapping the diversity matrix to the forecast errors. The proposed approach yielded comparable forecasting performance with the top-performing methods in the M4 competition.

As expected, the implementations of stacking in a cross-learning manner also come with their limitations. The first limitation is the requirement for a large, diverse time series dataset to enable meaningful training outcomes. This issue can be addressed by simulating series on the basis of some assumed data generating processes \citep{Talagala2018-meta} \citep[implemented using the \proglang{R} package \pkg{forecast},][]{rforecast}, or by generating time series with diverse and controllable characteristics \citep{Kang2020-rl} \citep[implemented in the \proglang{R} package \pkg{gratis},][]{rgratis}. Moreover, given the considerable literature on feature identification and feature engineering \citep[e.g.,][]{Wang2009-hs,Kang2017-wt,Lemke2010-wn,Montero-Manso2020-tq,Li2020-od}, the feature-based forecast combination methods naturally raise some issues yet to receive much research attention including how to design an appropriate feature pool in order to achieve the best out of such methods, and what is the best loss function for the meta-model.

It is also worth mentioning that many neural network models rely on a model combination strategy, namely ``ensembling'' \citep[see, e.g.,][a popular work in the machine learning context]{Caruana2004-en}, that is applied internally to improve the overall forecasting performance. Due to the weak learning process in deep learning models, the overall forecasting results heavily depend on the combination of each forecasting result. They diversify the individual forecast via (1) varying the training data, (2) varying the model pool, and (3) varying the evaluation metric. For example, the N-BEATS model \citep{Oreshkin2019-nb} utilized different strategies to diversify the forecasting results. For each forecasting horizon, individual models are trained with six window lengths. It also used three metrics sMAPE, MASE and MAPE to validate each model. In the end, a variety of models are used to make the median ensemble for results on the test set. One may refer to \citet{Ganaie2022-ed} for a general view of deep learning ensembles.

\subsection{Which forecasts should be combined?}
\label{sec:forecasts}

Including forecast methods with poor accuracy degrades the performance of the forecast combination. One prefers to exclude component forecasts that perform poorly and to combine only the top performers. In judgmental forecasting, \citet{Mannes2014-dl} highlighted the importance of the crowd's mean level of accuracy (expertise). They argued that the mean level of expertise sets a floor on the performance of combining. The gains in accuracy from selecting top-performing forecasts for combination have been investigated and confirmed by a stream of articles such as \citet{Budescu2015-tu} and \citet{Kourentzes2019-na}. \citet{Lichtendahl2020-ut} emphasized that the variance of accuracy across time series, which provides an indication of the accuracy risk, exerts a great influence on the performance of the combined forecasts. They suggested balancing the trade-offs between the average accuracy and the variance of accuracy when choosing component models from a set of available models.

Another key issue is diversity. Diversity among the individual forecasts is often recognized as one of the elements required for accurate forecast combination \citep{Batchelor1995-ps,Brown2005-aa,Thomson2019-al}. \citet{Atiya2020-ge} utilized the bias-variance decomposition of MSE to study the effects of forecast combinations and confirmed that an increase in diversity among the individual forecasts is responsible for the error reduction achieved in combined forecasts. Diversity among individual forecasts is frequently measured in terms of correlations among their forecast errors, with lower correlations indicating a higher degree of diversity. The distance of top-performing clusters introduced by \citet{Lemke2010-wn}, where a $k$-means clustering algorithm is applied to construct clusters, and a measure of coherence proposed by \citet{Thomson2019-al} are also considered as other measures to reflect the degree of diversity among forecasts.

In an analysis of a winner-take-all forecasting competition, \citet{Lichtendahl2013-ws} found that the optimal strategy for reporting forecasts is to exaggerate the forecasters' own private information and down-weight any common information. This exaggeration results in gains in the accuracy of the simple average by amplifying the diversity of the individual forecasts. The gains were confirmed by \citet{Grushka-Cockayne2017-dj}, who looked more closely at the impact of private-signal exaggeration on forecast combinations, which translates into averaging forecasts that are overfitted and overconfident.

Ideally, we would choose independent forecasts to amplify the diversity of the component forecasts when forming a combination. However, the available individual forecasts are often produced based on similar training, similar models and overlapping information sets, leading to highly positively correlated forecast errors. Including forecasts that have highly correlated forecast errors in a combination creates redundancy and may result in unstable weights, especially in the class of regression-based combinations (see Section~\ref{sec:linear_comb}). In this respect, using different types of forecasting models (e.g., statistical, machine learning, and judgmental), or different sources of information (e.g., exogenous variables), can help improve diversity \citep{Atiya2020-ge}. The results of the M4 competition reconfirmed the benefits of combinations of statistical and machine learning models \citep{Makridakis2020-hu}.

It is often suggested that a subset of individual forecasts be combined, rather than the full set of forecasts, as there are decreasing returns to adding additional forecasts \citep{Armstrong2001-sj,Zhou2002-cg,Hibon2005-ok,Geweke2011-xk,Diebold2019-ml,Lichtendahl2020-ut}. Simply put, \textit{many could be better than all}. In this regard, given a method pool with many forecasting models available, one can consider an additional step ahead of combining: subset selection. Instead of using all available forecasts in a combination, the step aims to eliminate some forecasts from the combination and select only a subset of the available forecasts.

The most common technique of subset selection is to include only the most accurate methods in the combination, discarding the worst-performing individual forecasts \citep[e.g.,][]{Granger2004-sw}. \citet{Mannes2014-dl} investigated the gains in accuracy from this \textit{select-crowd} strategy. \citet{Kourentzes2019-na} proposed a heuristic, where we exclude component forecasts that show a sharp drop in performance by using the outlier detection methods in boxplots. Their empirical results over four diverse datasets showed that this subset selection approach outperforms selecting a single forecast or combining all available forecasts. Nonetheless, the approach may suffer from a lack of diversity when formulating appropriate pools.

Early studies considering diversity used forecast encompassing tests for combining forecasts \citep[e.g.,][]{Kicsinbay2010-et,Costantini2010-hp}. The forecast encompassing literature ties in very closely with forecast combinations. Several forecast encompassing tests have been developed to test whether one forecast (or a set of forecasts) encompasses all information contained in another forecast (or another set of forecasts); see, e.g., \citet{Chong1986-ec} and \citet{Harvey1998-fe}. A classical argument suggests that when fixed weights are used (as in an average), only non-encompassed individual models are worth combining \citep{Diebold1989-fe}. However, \citet{Hendry2004-pf} provided a counter example in processes subject to location shifts where previously encompassed models may later dominate, while the earlier dominant model may later fail badly.

The diversity of an available forecast pool has occasionally been explicitly considered for subset selection. \citet{Cang2014-tp} proposed an optimal subset selection algorithm for forecast combinations based on mutual information, which takes account of diversity among different forecasts. More recently, \citet{Lichtendahl2020-ut} developed a subset selection approach comprising two screens: one screen for removing individual models that perform worse than the Na\"{i}ve2 benchmark, and another for excluding pairs of models with highly correlated forecast errors. In this way, both accuracy and diversity issues are addressed when forming a combination.

Subset selection techniques take advantage of allowing many forecasts to be considered when combining, reducing weight estimation errors and improving computational efficiency. However, subset selection has received scant attention in the context of forecast combinations, and it is mainly focused on trimming based on the principles of expertise. Therefore, automatic selection techniques considering both expertise and diversity merit further attention and development.

One approach is to note that subset selection is equivalent to assigning zero weights to some individual forecasts, which could be determined either statistically or judgmentally. \citet{Diebold2019-ml} focused on weights that solve a penalized estimation problem. Specifically, they proposed a two-step LASSO-based procedure that selects a subset of forecasts to combine in the first step, and shrinks the weights of the selected candidates toward equality. An alternative idea can be using a pre-set threshold to select individual models with weights greater than the threshold to join the subsequent combination; see, e.g., \citet{Zhou2002-cg,Wang2021-un}. Of course, there is no guarantee that the zero weight over the training period will also be zero over the forecast horizon. Hence, time-varying subset selection is certainly one solution to this problem and can be achieved by applying a pre-set threshold to forecast combinations with time-varying weights~\citep{Li2021-gk}.

\subsection{Forecast combination puzzle}
\label{sec:puzzle}

Despite the explosion of a variety of popular and sophisticated combination methods, empirical evidence and extensive simulations repeatedly show that the simple average with equal weights often outperforms more complicated weighting schemes. This somewhat surprising result has occupied a very large literature, including the early studies by \citet{Stock1998-np,Stock2003-sp,Stock2004-rq}, the series of Makridakis competitions \citep{Makridakis1982-hb,Makridakis2000-he,Makridakis2020-hu}, and also the more recent articles by \citet{Blanc2016-sn,Blanc2020-pg}, etc. \citet{Clemen1989-fb} surveyed the early combination studies and raised a variety of issues that remain to be addressed, one of which is ``What is the explanation for the robustness of the simple average of forecasts?'' In a recent study, \citet{Gastinger2021-ey} investigated the forecasting performance of a collection of combination methods on many time series from diverse sources and found that the winning combination methods differ for the different data sources, while the simple average strategies show, on average, more gains in improving accuracy than other more complex methods. \citet{Stock2004-rq} coined the term ``forecast combination puzzle'' for the phenomenon --- theoretically sophisticated weighting schemes should provide more benefits than the simple average from forecast combination, while empirically the simple average has been continuously found to dominate more complicated approaches to combining forecasts.

Most explanations of why simple averaging might dominate complex combinations in practice have centered on the errors that arise when estimating the combination weights. For example, \citet{Timmermann2006-en} noted that the success of simple combinations is due to the increased parameter estimation error with weighted combinations --- simple combination schemes do not require estimating combination parameters, such as weights based on forecast errors. \citet{Smith2009-wd} demonstrated that the simple average is expected to overshadow the weighted average in a situation where the weights are theoretically equivalent. The results from simulations and an empirical study showed the estimation cost of weighted averages when the optimal weights are close to equality, thus providing an empirical explanation of the puzzle. Later, \citet{Claeskens2016-pv} provided a theoretical explanation for these empirical results. Taking the estimation of ``optimal'' weights (see Section~\ref{sec:linear_comb}) into account, \citet{Claeskens2016-pv} considered random weights rather than fixed weights during the optimality derivation and showed that, in this case, the forecast combination may introduce biases in combinations of unbiased component forecasts and the variance of the forecast combination may be larger than in the fixed-weight case, such as the simple average. More recently, \citet{Chan2018-jl} proposed a framework to study the theoretical properties of forecast combinations. The proposed framework verified the estimation error explanation of the ``forecast combination puzzle'' and, more crucially, provided additional insights into the puzzle. Specifically, the mean squared forecast error (MSFE) can be considered as a variance estimator of the forecast errors which may not be consistent, leading to biased results with different weighting schemes based on a simple comparison of MSFE values. \citet{Blanc2020-pg} explained why, in practice, equal weights are often a good choice using the tradeoff between bias (reflecting the error resulting from underfitting training data when choosing equal weights) and variance (quantifying the error resulting from the uncertainty when estimating other weights).

Explaining the puzzle using estimation error requires a hypothesis that potential gains from the ``optimal'' combination are not too large so that estimation error overwhelms the gains. Special cases, such as where the covariance matrix of the forecast errors has equal variances on the diagonal, and all off-diagonal covariances are equal to a constant, are illustrated by \citet{Timmermann2006-en} and \citet{Hsiao2014-ug} to arrive at equivalence between the simple average and the ``optimal'' combination. \citet{Elliott2011-ab} characterized the potential bounds on the size of gains from the ``optimal'' weights over the equal weights and illustrated that these gains are often too small to balance estimation error, providing a supplementary explanation of the puzzle for the explanation of large estimation error.

Rather than focusing on the impact of combination weight estimation, \citet{Zischke2022-sv} instead explored the impact of sampling variability in forecast combinations. They demonstrated that, asymptotically, the sampling variability in the performance of the combination forecast is driven entirely by the variability arising from the estimation of the constituent models, and combination weight estimation imparts no bias or variability to the performance of forecast combinations, which lies in opposition to the finding of \citet{Claeskens2016-pv}. These findings imply that, when the combination weights are theoretically equivalent, there will be little performance difference between a sophisticated forecast combination and an equally weighted combination, providing new insights into the ``forecast combination puzzle''.

The examination and explanation of the ``forecast combination puzzle'' can provide decision makers with the following guidelines to identify which combination method to choose in specific forecasting problems.
\begin{itemize}
  \item Estimation errors are identified as ``finite-sample estimation effects'' in \citet{Smith2009-wd}, which suggests that an insufficiently small sample size may be unable to provide robust weight estimates. Thus, if one has access to limited historical data, the simple average or estimated weights with covariances between forecast errors being neglected are recommended. In addition, alternative simple combination operators such as trimmed and winsorized means can be adopted to eliminate extreme forecasts, and thus, offer more robust estimates than the simple average.
  \item Structural changes, which may cause different weight estimates in the training and evaluation samples, tend to impact sophisticated combination approaches more than the simple average. This case makes the simple average the better choice. The forecast combinations using changing weights can also be considered as a means to cope with structural changes, as suggested in \citet{Diebold1987-go} and \citet{Deutsch1994-ob}.
  \item If one has access to many component forecasts, the PCR and the clustering strategy (for details, see Section~\ref{sec:linear_comb}) might be useful to diminish estimation errors and solve the multicollinearity problem by reducing the number of parameters need to be estimated.
  \item Involving time series features (see Section~\ref{sec:comb_learn}) and diverse individual forecasts (see Section~\ref{sec:forecasts}) in the process of weight estimation can enlarge the gains of forecast combinations, providing a possible way to untangle the ``forecast combination puzzle''.
\end{itemize}
In summary, forecasters are encouraged to analyze the data prior to identifying the combination strategy and to choose combination rules tailored to specific forecasting problems.

\section{Probabilistic forecast combinations}
\label{sec:probabilistic}

\subsection{Probabilistic forecasts}

In recent years, probabilistic forecasts have received increasing attention. For example, the recent Makridakis competitions, the M4 and the M5 Uncertainty \citep{Makridakis2020-lz} competitions, encouraged participants to provide probabilistic forecasts of different types as well as point forecasts. Probabilistic forecasts are appealing for enabling optimal decision-making with better understanding of uncertainties and the resulting risks. A brief survey of extensive applications of probabilistic forecasting was offered by \citet{Gneiting2014-tz}.

Probabilistic forecasts can be reported in various forms including density forecasts, distribution forecasts, quantiles, and prediction intervals, and how to combine them can vary. For example, although a quantile forecast is the inverse of the corresponding forecast represented by the cumulative distribution function, the combined quantile forecast and the combined probability forecast may not be equivalent. Simple examples of averaging quantiles and probabilities with equal weights are provided by \citet{Lichtendahl2013-rt}.

Interval forecasts form a crucial special case and are often constructed using quantile forecasts where the endpoints are specific quantiles of a forecast distribution. For example, the lower and upper endpoints of a central $(1-\alpha)\times 100\%$ prediction interval can be defined via the quantiles at levels $\alpha/2$ and $1-\alpha/2$.

As with point forecasts, combining multiple probabilistic forecasts allows for diverse information sets and different types of forecasting models, as well as the mitigation of potential misspecifications derived from a single model. Empirical studies suggest that the relative performance of different models often varies over time due to structural instabilities in the unknown data generating process \citep[e.g.,][]{Billio2013-sg}. Thus, there has been a growing interest in bringing together multiple probabilistic forecasts to produce combined forecasts that integrate information from separate sources.

\subsection{Scoring rules}
\label{sec:issues}

Decision makers mainly focus on accuracy when combining point forecasts, while other measures such as calibration and sharpness need to be considered when working with combinations of probabilistic forecasts \citep{Gneiting2007-fr,Gneiting2007-ij,Lahiri2015-qq}. \textbf{Calibration} concerns the statistical consistency between the probabilistic forecasts and the corresponding realizations, thus serving as a joint property of forecasts and observations. In practice, a probability integral transform (PIT) histogram is commonly employed informally as a diagnostic tool to assess the calibration of probability forecasts regardless of whether they are continuous \citep{Dawid1984-vp,Diebold1998-cr} or discrete \citep{Gneiting2013-hl}: A uniform histogram indicates a probabilistically calibrated forecast. \textbf{Sharpness} refers to the concentration of probabilistic forecasts, and thus serves as a property of the forecasts only; the sharper a forecast is, the better it is. Sharpness is easily comprehended when considering prediction intervals: the sharper the forecasts, the narrower the intervals. In the case of probability forecasts, sharpness can be assessed in terms of the width of central prediction intervals. For more thorough definitions and diagnostic tools of calibration and sharpness, we refer to \citet{Gneiting2014-tz}.

According to \citet{Gneiting2007-fr}, the intent of probabilistic forecasting is to \textit{maximize the sharpness of the forecast distributions subject to calibration} based on the available information set. In this light, scoring rules that reward both calibration and sharpness are appealing in the sense of providing summary measures for the quality of probabilistic forecasts, with a higher score indicating a better forecast. For a probabilistic forecast $F$, a scoring rule is proper if it satisfies the condition that the expected score for an observation drawn from distribution $G$ is maximized when $F=G$. It is strictly proper if the maximum is unique. \citet{Gneiting2007-ij} provides an excellent review and discussion on a diverse collection of proper scoring rules for probabilistic forecasts.

The schemes for combining multiple probabilistic forecasts have evolved from a simple distribution mixture to more sophisticated combinations accounting for correlations between distributions. Which type of strategy one might choose to use depends largely on the computational burden, and the overall performance of the combined forecasts with regard to accuracy, calibration, and sharpness.

\subsection{Linear pooling}
\label{sec:linear_pooling}

Probability forecasts strive to predict the probability distribution of quantities or events of interest. In line with the notations in previous sections, here we consider $N$ individual forecasts specified as cumulative probability distributions of a random variable $Y$ at time $T+h$, denoted $F_{i}(y_{T+h}|I_{T})$, $i=1,\dots,N$, using the information available up to time $T$, $I_{T}$. One popular approach is to directly take a mixture distribution of these $N$ individual probability forecasts with estimated weights, neglecting correlations between these individual components. This approach is commonly referred to as the ``linear opinion pool'' in the literature on combining experts' subjective probability distributions, dating back at least to \citet{Stone1961-zd}. The linear pool of probability forecasts is defined as the finite mixture
\begin{align}
  \label{eq:linear_pool}
  \tilde{F}(y_{T+h}|I_{T}) = \sum_{i=1}^{N} w_{T+h|T,i} F_{i}(y_{T+h}|I_{T}),
\end{align}
where $w_{T+h|T,i}$ is the weight assigned to the $i$th probability forecast. These weights are often set to be non-negative and sum to one to guarantee that the pooled forecast preserves properties of both non-negativity and integrating to one. The pooled probability forecast satisfies numerous properties such as the \textit{unanimity} property (if all individual forecasters agree on a probability then the pooled forecast agrees also); see \citet{Clemen1999-mh} for more details.

Linear pooling of probability forecasts allows us to accommodate skewness and kurtosis (fat tails), and also multi-modality, even under normal distributions of individual forecasts; see \citet{Wallis2005-yf} and \citet{Hall2007-lh} for further discussion on this point.

Define $\mu_{i}$ and $\sigma_{i}^{2}$ as the mean and variance of the $i$th component forecast distribution and drop the time and horizon subscripts for simplicity. Then the linear combined probability forecast has the mean and variance,
\begin{align}
   & \tilde{\mu} = \sum_{i=1}^{N} w_{i} \mu_{i}, \label{eq:mean_linear_pooling}  \\
  \text{and}\qquad
   & \tilde{\sigma}^{2} = \sum_{i=1}^{N} w_{i} \sigma_{i}^{2} + \sum_{i=1}^{N} w_{i} \left(\mu_{i}-\tilde{\mu}\right)^{2}. \label{eq:variance_linear_pooling}
\end{align}
Note that the mean of the combination distribution is equivalent to the linear combination of the individual means. Thus, the associated combination point forecast is consistent with the linear combination point forecast.

However, the variance of the combination distribution is larger than the linear combination of the individual variances when the individual means differ. Consequently, the common strategy of seeking diverse forecasts may harm the probabilistic forecast, while helping the point forecast; see \citet{Ranjan2010-jl} for a theoretical illustration and simulation study. Simply put, as the diversity among individual probability forecasts increases, the mixed forecast will lose sharpness and may become under-confident because of the spread driven by the disagreement between the individual probability forecasts \citep{Hora2004-fz,Wallis2005-yf,Ranjan2010-jl}.

Even in the ideal case in which individual forecasts are well calibrated, the resulting linear pooling combination may be poorly calibrated. Theoretical aspects of this finding and properties of linear pools of probability forecasts have been further studied in \citet{Hora2004-fz}, \citet{Ranjan2010-jl}, and \citet{Lichtendahl2013-rt}.

On the other hand, \citet{Hora2004-fz} demonstrated, both from theoretical and empirical aspects, that linear pooling may work to provide better calibrated forecasts than the individual distributions when individual forecasts tend to be overconfident. This finding helps to account for the success of linear pooling in varied applications. \citet{Jose2014-uh} highlighted that if the experts are overconfident but have low diversity, the linear pool may remain overconfident. \citet{Lichtendahl2013-rt} identified three factors that manipulate the calibration of the probability forecast derived from linear pooling: (i)~the number of constituent forecasts, (ii)~the degree to which the constituent forecasts are overconfident, and (iii)~the degree of the constituents' disagreement on the location (e.g., mean) of the distribution.

In principle, probability forecasts can be recalibrated before or after the pooling to correct for miscalibration \citep{Turner2014-za}. However, it is challenging to appraise the degree of miscalibration, which may vary considerably among different forecasts and over time, and therefore to recalibrate accordingly. Some effort has been directed toward the development of alternative combination methods to address the calibration issue. For example, \citet{Jose2014-uh} suggested the ``trimmed opinion pool'', which trims away some individual forecasts from the ``linear opinion pool'' before mixing the component forecasts. Specifically, exterior trimming that trims away forecasts with low or high means values serves as a way to address under-confidence by decreasing the variance. Conversely, interior trimming that trims away forecasts with moderate means is suggested to mitigate overconfidence via increasing the variance. The improvement in forecasting performance offered by trimming was confirmed by \citet{Grushka-Cockayne2017-dj} at a more foundational level.

Some researchers prefer nonlinear alternatives, including a generalized linear pool, the spread-adjusted linear pool, and the beta-transformed linear pool, in terms of delivering better calibrated combined probability forecasts; these are discussed in Section~\ref{sec:nonlinear_pooling}. Instead of mixing probability forecasts mentioned above, \citet{Lichtendahl2013-rt} recommended averaging quantile forecasts (see Section~\ref{sec:quantile_comb}) based on the supportive results both theoretically and empirically.

The key practical issue determining the success (or failure) of linear pooling is how the weights for the individual probability forecasts in the finite mixture should be estimated. As with point forecast combinations, equal weights are worthy of consideration, while determining optimal weights is particularly challenging in the case of having access to probability forecasts with limited historical data.

Linear pooling with equal weights is easy to understand and implement, commonly yielding robust and stable outcomes. For reviews, see, e.g., \citet{Wallis2005-yf} and \citet{OHagan2006-jk}. A leading example is the survey of professional forecasters (SPF) in the US, which publishes mixed probability forecasts (in the form of histograms) for inflation and GDP growth using equal weights. As the experience of combining point forecasts has taught us, the equally weighted approach often turns out to be hard to beat. An important reason is that it avoids parameter estimation error that often exists in weighted approaches; see Section~\ref{sec:puzzle} for more details and illustrations.

Motivated by the ``optimal'' weights obtained in point forecast combinations by minimizing the MSE loss (see Section~\ref{sec:linear_comb}), \citet{Hall2007-lh} proposed obtaining the set of weights by minimizing the Kullback-Leibler information criterion (KLIC) distance between the combined probability forecast density $\tilde{f}(y_{\tau+h}|I_{\tau})$ and the true (but unknown) probability density $f(y_{\tau+h})$, $\tau=1,\dots,T$. The KLIC distance is defined as
\begin{align*}
  \mathrm{KLIC} = \int f(y_{\tau+h}) \log \left\{\frac{f(y_{\tau+h})}{\tilde{f}(y_{\tau+h}|I_{\tau})}\right\} \mathrm{d} y_{\tau+h}
                =E\left[\log f(y_{\tau+h})-\log \tilde{f}(y_{\tau+h}|I_{\tau})\right].
\end{align*}

Under the asymptotic assumption that the number of observations $T$ grows to infinity, the problem of minimizing the KLIC distance reduces to the maximization of the average logarithmic score of the combined probability forecast. Therefore, the optimal weight vector $\bm{w}_{T+h|T}$ is given by
\begin{align}
  \label{eq:weight_klic}
  \bm{w}_{T+h|T} = \underset{\bm{w}}{\operatorname{argmax}} \frac{1}{T-h} \sum_{t=1}^{T-h} \log \tilde{f}(y_{t+h}|I_{t}),
\end{align}
where $\bm{w}_{T+h|T}=\left(w_{T+h|T, 1}, \dots, w_{T+h|T, N}\right)^{\prime}$. The use of the logarithmic scoring rule eliminates the need to estimate the unknown true probability distribution, and therefore simplifies the weight estimation for the component forecasts. This was followed by \citet{Pauwels2016-ci} documenting the properties of the optimal weights in Equation~\eqref{eq:weight_klic}, centering on the asymptotic assumption used by \citet{Hall2007-lh}. Their simulations and empirical results indicated that the combination with optimal weights is inferior for small $T$, while it is valid in minimizing the KLIC distance when $T$ is sufficiently large. Therefore, a sufficient training sample is recommended when solving the optimization problem.

Following in the footsteps of \citet{Hall2007-lh}, many extensions and refinements of the combination strategy have been suggested. \citet{Conflitti2015-fq} devised a simple iterative algorithm to compute the optimal weights in Equation~\eqref{eq:weight_klic}. The algorithm scales well with the dimension $N$, and hence enables the combination of many individual probability forecasts. \citet{Geweke2011-xk} provided a Bayesian perspective on an optimal linear pool, and provided a theoretical justification for the use of optimal weights. \citet{Li2021-gk} conducted time-varying weights based on time-varying features from historical information, where the weights in the forecast combination were estimated via Bayesian logarithmic predictive scores. \citet{Jore2010-yi} put forward an exponential weighting scheme based on the recursive weights constructed using the relative past performance of each individual probability forecast in terms of the logarithmic score. In contrast to the optimal opinion pool based on the weights in Equation~\eqref{eq:weight_klic}, in this case, the logarithmic score of the combined probability forecast is not necessarily maximized. The logarithmic scoring rule is appealing as it intuitively assigns a higher weight to a component forecast that better fits the realized value. On the other hand, forecast combinations with weights optimized by minimizing the continuously ranked probability score \citep[CRPS,][]{Gneiting2007-ij}, which is a strictly proper scoring rule for distribution forecasts, have been considered in some research, see, e.g, \citet{Raftery2005-vu}, \citet{Thorey2017-on}, and \citet{Thorey2018-en}.

Furthermore, some special treatments were given to accommodate probability forecast combinations in applications such as energy forecasting, retail forecasting, and economic forecasting. For instance, \citet{Opschoor2017-yu} extended the idea of optimal combinations but estimated optimal weights by either maximizing the censored likelihood scoring rule \citep{Diks2011-gj} or minimizing a weighted version of the CRPS, allowing forecasters to limit themselves to a specific region of the target distribution. For example, we are more likely to be interested in avoiding out-of-stocks when working with retail forecasting. The tail of the distribution is also the main feature of interest when measuring downside risk in equity markets. Additionally, \citet{Zischke2022-sv} showed that when forecasting during times of high volatility, forecast combinations produced by optimizing according to the censored likelihood scoring rule always lead to a better out-of-sample performance than ``optimal'' forecast combinations with weights optimized using the logarithmic score, which lends support to the use of a scoring rule that prioritizes accurate forecasts in a specific region. \citet{Diebold2022-rm} instead constructed regularized mixtures of density forecasts using a variety of objectives and regularization penalties. The optimal regularization tends to spread probability mass from the center into both tails of the distribution, correcting for overconfidence and adjusting kurtosis. Besides, \citet{Pauwels2020-zl} proposed an approach to computing the optimal weights by maximizing the average logarithmic score subject to additional higher moments restrictions. Through constrained optimization, the combined probability forecast can preserve specific characteristics of the distribution, such as fat tails or asymmetry. \citet{Martin2021-yi} looked at mode misspecification, and showed via simulation and empirical results that score-specific optimization of linear pooling weights does not always achieve improvements in forecasting accuracy.

\subsection{Bayesian model averaging}
\label{sec:bma}

Bayesian model averaging (BMA) provides an alternative means of mixing individual probability forecasts with respect to their posterior model probabilities. BMA offers a conceptually elegant and logically coherent solution to the issue of accounting for model uncertainty \citep[see, e.g.,][]{Leamer1978-sp,Draper1995-cd,Raftery1997-ij,Garratt2003-kh}. Under this approach, the posterior probability forecast is computed by mixing a set of individual probability forecasts distributions, $F_{i}(y_{T+h}|I_{T})=F(y_{T+h}|I_{T},M_{i})$, from model $M_{i}$, and can be given as
\begin{align}
  \label{eq:BMA}
  \tilde{F}(y_{T+h}|I_{T}) = \sum_{i=1}^{N} P(M_{i}|I_{T})F(y_{T+h}|I_{T},M_{i}),
\end{align}
where $P(M_{i}|I_{T})$ is the posterior probability of model $M_{i}$. The decision makers update the prior probability of model $M_{i}$ being the true model, $P(M_{i})$, via Bayes' Theorem to compute the posterior probability
\begin{align}
  \label{eq:post_prob}
  P(M_{i}|I_{T}) & = \frac{P(M_{i})P(I_{T}|M_{i})}{\sum_{i=1}^{N} P(M_{i})P(I_{T}|M_{i})},
\end{align}
where
\begin{align}
  \label{eq:marginal_likelihood}
  P(I_{T}|M_{i}) = \int_{\bm{\theta}_{i}} P\left(\bm{\theta}_{i} | M_{i}\right) P\left(I_{T} | M_{i}, \bm{\theta}_{i}\right) d \bm{\theta}_{i}
\end{align}
is the marginal likelihood of model $M_{i}$, $P\left(\bm{\theta}_{i} | M_{i}\right)$ is the prior on the unknown parameters $\bm{\theta}_{i}$ conditional on model $M_{i}$, and $P\left(I_{T} | M_{i}, \bm{\theta}_{i}\right)$ is the likelihood function of model $M_{i}$. See, e.g., \citet{Koop-2003bayesian} for textbook illustrations of BMA.

BMA in Equation~\eqref{eq:BMA} can be viewed as a form of linear pooling of individual probability forecasts \eqref{eq:linear_pool}, weighted by their posterior model probabilities given in Equation~\eqref{eq:post_prob}. Note that the weights characterized by posterior probabilities do not account for correlations among individual probability forecasts. The approach provides a general way to deal with model uncertainty and does not necessarily require the use of conjugate families of distributions. The BMA procedure is consistent in the sense that the posterior probability in Equation~\eqref{eq:post_prob} indicates the probability that model $M_{i}$ is the best under the KLIC measure distance and shows how well the model fits the observations \citep{Fernandez-Villaverde2004-pz,Raftery2005-vu,Wright2008-bs}.

While theoretically attractive, BMA suffers from three major challenges when implemented in practice. One is how to properly specify the model space of interest to avoid model incompleteness. It is often impractical to cover the complete set of models when the number of possible models is large or their structures are complex. This difficulty can mostly be resolved via the selection of a subset of models that are supported by the data or through stochastic search algorithms over the model space; see \citet{Hoeting1999-qn} and \citet{Koop2003-nl} for more details on model search strategies. A second well-known challenge relates to the elicitation of two types of priors (on parameters and models) for many models of interest \citep{Moral-Benito2015-zh,Aastveit2019-lf}. Another practical concern lies in the computation of the integrals in Equation~\eqref{eq:marginal_likelihood}. The integrals that are required for the derivation of the marginal likelihood may be analytically intractable in many cases, except for the generalized linear regression models with conjugate priors. The Laplace method as well as the Markov chain Monte Carlo (MCMC) methods are therefore frequently used to provide an excellent approximation to $P(I_{T}|M_{i})$; see, e.g., \citet{Hoeting1999-qn} and \citet{Bassetti2020-uh} for discussions of these approximations.

One drawback of the BMA approach is the implicit assumption that the true model is included in the model space to be considered \citep{Wright2008-bs}. Under this assumption, when the sample size tends to infinity, the posterior probabilities converge to zero, except for one which converges to unity. Thus, BMA reduces to model selection for large sample size, with the best model (which is the true model if that exists, but is still well-defined if none of the models is in fact true) receiving a weight very close to one; see \citet{Geweke2010-zn} for an empirical demonstration. In this regard, the combined forecast derived from BMA may be misspecified when the model space is incomplete (i.e. all models under consideration are incorrect), arising the issue of model incompleteness. Recently, \citet{Yao2018-st} took the idea of stacking from the literature on point forecast combinations (see Section~\ref{sec:comb_learn}) and generalized it to the combinations of forecast distributions in the Bayesian setting, which can essentially be regarded as a minor tweak on BMA. However, as the critique given at the end of \citet{Yao2018-st} says, averaging distribution functions may be inferior to averaging quantiles (see Section~\ref{sec:quantile_comb}), especially when the combination problem is more like an information aggregation problem rather than a BMA problem, and BMA (or minor tweaks on it) does not seem like the right framework since we are almost always in a world where there is no true model.

In contrast, optimal weights as defined in Equation~\eqref{eq:weight_klic} do not suffer from the issue of model incompleteness because the weights need not converge to zero or unity regardless of whether the component models are correct or not, which allows for a convex combination (rather than a selection) of the individual probability forecasts distributions. In a binary-event context, \citet{Lichtendahl2022-bi} introduced a new class of Bayesian combinations in which stacking is used to form the approach by aggregating the probabilities provided by the experts or models. But it should not be confused with BMA and the approach developed by \citet{Yao2018-st}, since it does not have to assume, as BMA does, that one of the models being combined is the true model, and its setting is information aggregation rather than model selection. \citet{Lichtendahl2022-bi} showed that extremizing \citep[i.e. shifting the average probability closer to its nearest extreme, see, e.g.,][]{Satopaa2016-mo} is not always appropriate when combining binary-event forecasts.

The other drawback of the BMA approach may be related to the fixed probabilities assigned to component models, as documented in \citet{Aastveit2019-lf}. The uncertainty of the weights is ignored in this case, leading to unstable combined forecasts in a forecasting environment characterized by large instability and structural changes in the forecast performance of the individual models. Thus, it is plausible to let the pooling weights evolve over time. \citet{Raftery2010-qe} developed a model combination strategy for doing dynamic model averaging (DMA) which allows for the forecasting model as well as the coefficients in each model to evolve over time. Considering multiple models, the goal of DMA is to calculate the probabilities that the process is governed by model $M_{i}$ for $i=1,\dots,N$ at time $T+1$, given the information available up to time $T$, and average forecasts across individual models using these probabilities. When the forecasting model and model parameters do not change, DMA reduces to a recursive implementation of standard BMA. The strategy advocated by \citet{Raftery2010-qe} can also be used for dynamic model selection (DMS), where a single model with the highest probability is selected and used to forecast. Note that these calculated probabilities will vary over time and, thus, different forecasting models hold at each point in time. Such specifications are of particular interest in economics, see, e.g., \citet{Koop2012-dma} and \citet{Del_Negro2016-ab} for notable macroeconomic applications. One contribution of \citet{Raftery2010-qe} is that a forgetting factor is used to develop a computationally efficient recursive algorithm that allows for fast calculation of the required probabilities when model uncertainty and the number of models considered are large.

\subsection{Nonlinear pooling}
\label{sec:nonlinear_pooling}

Despite their simplicity and popularity, the classical linear pooling methods have several shortcomings, such as the calibration problem discussed previously. A linear pooling of probability forecasts increases the variance of the forecasts and may result in a suboptimal solution, lacking both calibration and sharpness. To address these shortcomings, several nonlinear alternatives to linear pooling methods have been developed for recalibration purposes.

Motivated by the seminal work of \citet{Dawid1995-jj}, \citet{Gneiting2013-hl} developed the generalized linear pool (GLP) to incorporate a parametric family of combination formulas. Let $F_{i}(y_{T+h}|I_{T})$ denote the cdf of the probability forecast ($i=1,\dots,N$), and $\tilde{F}(y_{T+h}|I_{T})$ denote the cdf of the combined forecast. The generalized pooling scheme takes the following form
\begin{align*}
  \tilde{F}(y_{T+h}|I_{T}) = g^{-1}\bigg(\sum_{i=1}^{N} w_{T+h|T,i} g\big(F_{i}(y_{T+h}|I_{T})\big)\bigg),
\end{align*}
where $w_{T+h|T,1},\dots,w_{T+h|T,N}$ are nonnegative weights that sum to one, and $g$ denotes a continuous and strictly monotonic function with the inverse $g^{-1}$. The linear, harmonic and logarithmic (geometric) pools become special cases of the GLP for $g(x)=x$, $g(x)=1/x$ and $g(x)=\log(x)$, respectively. \citet{Gneiting2013-hl} highlighted that the generalized pooling strategy may fail to be sufficiently flexibly dispersive for calibration.

As a result, they also proposed the spread-adjusted linear pool (SLP) to allow one to address the calibration problem. Define $F_{i}^{0}$ and corresponding density $f_{i}^{0}$ via $F_{i}(y_{T+h}|I_{T})=F_{i}^{0}(y_{T+h}-\eta_{i}|I_{T})$ and $f_{i}(y_{T+h}|I_{T})=f_{i}^{0}(y_{T+h}-\eta_{i}|I_{T})$, where $\eta_{i}$ is the unique median of $F_{i}(y_{T+h}|I_{T})$. Then the SLP has the combined cdf and the corresponding density,
\begin{align*}
   & \tilde{F}(y_{T+h}|I_{T})=\sum_{i=1}^{N} w_{T+h|T,i} F_{i}^{0}\left(\frac{y_{T+h}-\eta_{i}}{c}\bigg|I_{T}\right) \quad \text { and } \\
   & \tilde{f}(y_{T+h}|I_{T})=\frac{1}{c} \sum_{i=1}^{N} w_{T+h|T,i} f_{i}^{0}\left(\frac{y_{T+h}-\eta_{i}}{c}\bigg|I_{T}\right),
\end{align*}
respectively, where $w_{T+h|T,1},\dots,w_{T+h|T,N}$ are nonnegative weights with $\sum_{i=1}^{N}w_{T+h|T,i}=1$, and $c$ is a strictly positive spread adjustment parameter. The traditional linear pool arises as a special case for $c = 1$. A value of $c < 1$ is suggested for neutrally confident or underconfident component forecasts, while a value $c \geq 1$ is suggested for overconfident components. Moreover, one can introduce spread adjustment parameters varying with the components in case the degrees of miscalibration of the components differ substantially.

The cumulative beta distribution is widely employed for recalibration because of the flexibility of its shape \citep[see, e.g.,][]{Graham1996-qc}. \citet{Ranjan2010-jl} introduced a beta-transformed linear pool (BLP) that merges the traditional linear pool with a beta transform to achieve calibration. The BLP takes the form
\begin{align*}
  \tilde{F}(y_{T+h}|I_{T}) = B_{\alpha, \beta}\bigg(\sum_{i=1}^{N} w_{T+h|T,i} F_{i}(y_{T+h}|I_{T})\bigg),
\end{align*}
where $w_{T+h|T,1},\dots,w_{T+h|T,N}$ are nonnegative weights that sum to one, and $B_{\alpha, \beta}$ is the cdf of the beta distribution with the shape parameters $\alpha > 0$ and $\beta > 0$. Full generality of the BLP enables an asymmetric beta-transformation on the basis of the linear pooling of probability forecasts. In its most simplistic case, the BLP approach nests the traditional linear pool, under the restriction $\alpha = \beta = 1$. The beta-transformation tunes up a linear pooled probability forecast if it is larger than $0.5$ and tunes it down otherwise when imposing the constraint $\alpha = \beta \geq 1$. The approach can be used to combine probability forecasts from both calibrated and uncalibrated sources. The estimates of the beta-transformation along with the mixture weights for linear pooling can be obtained by maximum likelihood, as suggested by \citet{Ranjan2010-jl}. Recent work by \citet{Lahiri2015-qq} demonstrated the superiority of the BLP approach, based on identifying the most valuable individual forecasts by a Welch-type test, over the equally weighted approach with respect to calibration and sharpness.

To achieve improved calibration properties, \citet{Bassetti2018-qr} proposed a Bayesian nonparametric approach, which is based on Gibbs and slice sampling, to realize the calibration and combination of probability forecasts by introducing an additional beta mixture in the BLP method. The resulting predictive cdf is
\begin{align*}
  \tilde{F}(y_{T+h}|I_{T}) = \sum_{k=1}^{K} \omega_{k} B_{\alpha_{k}, \beta_{k}}\bigg(\sum_{i=1}^{N} w_{T+h|T,ki} F_{i}(y_{T+h}|I_{T})\bigg),
\end{align*}
where $\omega_{1}, \dots, \omega_{K}$ denote beta mixture weights. The proposed approach enables one to treat the parameter $K$ as bounded or unbounded and it reduces to the BLP for $K=1$. The Bayesian inference approach achieved a compromise between parsimony and flexibility and produced well calibrated and accurate forecasts in their simulations and the empirical examples, outperforming the linear pool substantially.

The essence of these nonlinear pooling methods is to perform various transformations, that may be nonlinear, to either the component forecasts or the linearly pooled forecasts, in order to restore calibration and sharpness. \citet{Kapetanios2015-bb} generalized the literature by incorporating the dependence of the mixture weights on the variable one is trying to forecast, allowing the weights themselves to introduce the nonlinearities and thus leading to outcome-dependent density pooling. Clearly, the forecast performance of nonlinear pooling approaches largely depends on diverse factors, including the features of the target data, mixture component models, and training periods, and thereby deserves further research. This is in agreement with \citet{Baran2018-nm} who investigated the performance of state-of-the-art forecast combination methods through case studies and found no substantial differences in forecast performance between the simple linear pool and the theoretically superior but cumbersome nonlinear pooling approaches.

\subsection{Meteorological ensembles}
\label{sec:weather}

The term ``combination'' and ``ensemble'' are often used interchangeably in the literature on forecast combinations. However, ``ensemble'' was originally developed in the meteorological literature in a distinct way from the combinations of probabilistic forecasts that we have discussed so far.

Instead of combining multiple probabilistic forecasts available for forecasters, as with the approaches reviewed in preceding sections, an ensemble weather forecast is constructed from a set of point forecasts of the same weather quantity of interest, based on perturbed initial atmospheric states \citep[e.g.,][]{Maqsood2004-pe,Gneiting2005-yn} and/or different model formulations \citep[e.g.,][]{Buizza1999-st,Buizza2005-wf}. In this light, two major sources of forecast uncertainty, initial condition uncertainty resulting from the chaotic nature of the atmosphere, and model uncertainty arising from imperfect numerical models, are addressed \citep{Lorenz1963-yn,Weigel2008-vy,Baran2014-tm}. This enables a measure of uncertainty to be attached and makes an ensemble weather forecast more valuable than a single ``deterministic'' forecast, providing an inherently probabilistic assessment.

A meteorological ensemble forecast is a probabilistic forecast in the sense described here, assuming that there is no inherent uncertainty other than that contained in the initial conditions and the model formulation. In contrast, most statistical forecasting methods include an important additional source of uncertainty due to random noise innovations, but do not usually include uncertainty due to initial conditions. The distinction is important enough, and the literature sufficiently distinct, that we have chosen to discuss meteorological ensemble forecasts in this separate section.

It has been demonstrated that the raw meteorological ensemble forecasts typically present systematic errors regarding bias \citep{Atger2003-lx,Mass2003-bc} and dispersion \citep{Buizza2005-wf,Sloughter2010-ae}, with a tendency for the truth frequently falling outside of the range of the ensemble. Various statistical postprocessing methods have been introduced accordingly, with the aim of improving the forecast quality, to correct these errors by estimating representable relationships between the response variable of interest and predictors. Most postprocessing methods can be categorized into two groups: parametric approaches with distribution-based assumptions, such as ensemble model output statistics \citep[EMOS,][]{Gneiting2005-ua} models and BMA \citep{Raftery2005-vu}, and nonparametric approaches with distribution-free assumptions, such as the analog-based method \citep[e.g.,][]{Delle_Monache2013-os} and the quantile regression forest \citep{Taillardat2019-ni}. See \citet{Vannitsem2021-bg} for a recent review of statistical postprocessing methods as well as their potential and challenges.

Recently, the community of weather forecasting is starting to explore the potentials of machine learning techniques, especially in the context of ensemble forecasting, in the sense of including arbitrary predictors and accounting for nonlinear dynamics of the Earth system that are not captured by existing numerical models \citep{Dueben2021-fe}. One use of machine learning techniques is to complement ensemble NWP \citep[numerical weather prediction, see, e.g.,][for a summary of its revolution]{Bauer2015-dg,Benjamin2019-zp} forecasts using an additive postprocessing step for correction of ensemble bias and spread \citep{Rasp2018-zu,Scher2018-dm,Gronquist2021-no}. Machine learning techniques, such as neural networks, have also been used as data-driven forecast tools, an alternative to NWP models based on the physical laws governing the atmosphere, to generate base forecasts. These techniques lead to improved computational efficiency in creating ensemble forecasts with much larger ensemble sizes~\citep{Dueben2018-ln,Scher2018-of,Rasp2021-az,Scher2021-ee}.

\subsection{Combinations constructed via Bayes' Theorem}
\label{sec:bayesian_comb}

Pooling approaches, elaborated in Sections~\ref{sec:linear_pooling}--\ref{sec:nonlinear_pooling}, pool/mix multiple probability forecasts with equal weights, weights evaluated using various scoring rules, or posterior probabilities sequentially updated via Bayes' Theorem. They inherently neglect correlations among the component probability forecasts. Forecasts derived from different sources, nevertheless, are likely to share the same data, overlapping information, similar forecasting models, and common training processes. Thus, some sort of dependence among individual probability forecasts is extremely likely, and such dependence can have a serious impact on the aggregated distributions. In this section, we review an alternative class of combination techniques, in which dependence among component probability forecasts can be incorporated. The major feature, that makes this class of combinations difficult, lies in how to model the dependence among individual distributions in order to achieve good performance.

The extensive literature on probability forecast combinations considering correlations among individual distributions has, for the most part, been driven from a foundational Bayesian perspective and originated in agent opinion analysis theory, free from the time series context, dating back at least to the pioneering work of \citet{Winkler1968-uw}. We remark that in pooling techniques, the contribution of each individual probability forecast to the final aggregated probability forecast is measured explicitly via weights, whereas it is not specified by a specific form in the Bayesian combination techniques we discuss in this section.

Early work in the Bayesian vein focused on a Bayesian paradigm developed by \citet{Morris1974-yh,Morris1977-xl} in which a decision maker views available probability forecasts from various sources simply as data, and updates his/her prior distribution by means of Bayes' Theorem. At time $T$, the decision maker aims to forecast $y_{T+h}$ and receives current $h$-step-ahead probability forecasts $\mathcal{H}_{T+h} = \left\{f_{1}(y_{T+h}|I_{T}), \dots, f_{N}(y_{T+h}|I_{T})\right\}$ from the set of models. The posterior probability forecast of $y_{T+h}$ is then
\begin{align}
  \label{eq:bayes_theorem}
  \tilde{f}\left(y_{T+h} | I_{T}, \mathcal{H}_{T+h}\right) \propto p\left(y_{T+h} | I_{T}\right) f_{N}\left(\mathcal{H}_{T+h} | y_{T+h}, I_{T}\right),
\end{align}
where $p\left(y_{T+h} | I_{T}\right)$ denotes the decision maker's prior probability for $y_{T+h}$ given the available information $I_{T}$, and $f_{N}\left(\mathcal{H}_{T+h} | y_{T+h}, I_{T}\right)$ denotes the joint likelihood function derived from the individual distributions.

The problem of eliciting the posterior probability forecast in Equation~\eqref{eq:bayes_theorem} is therefore broken down into the problem of specifying the prior distribution and assessing the form of the joint distribution, or likelihood, derived from the component probability forecasts. A flat (possibly improper) prior is often considered in the literature because: (i)~it is reasonable to assume that everything the decision maker knows is integrated into the individual distributions; and (ii)~if not, the extra knowledge from the decision maker can be incorporated in the likelihood as an additional individual distribution; see, e.g., \citet{Winkler1968-uw}, \citet{Clemen1993-ty}, \citet{Clemen1985-kx}, and \citet{Jouini1996-fe}. Thus, the application of Bayes' Theorem presents the most taxing difficulties in delicately specifying the likelihood function, which requires consideration of the bias and precision of the individual distributions as well as their dependence \citep{Hall2007-lh}.

One line of research has considered specifying the likelihood as a joint distribution of forecast errors, and supported the use of the correlation between individuals' forecast errors in an effort to represent the dependence among individual distributions. Emphasis has been placed on making the likelihood computation tractable by adopting certain distributional assumptions. For example, \citet{Winkler1981-bn} assessed the likelihood as a multivariate normal distribution. Restricting the focus to individual models with unbiased forecasts, he derived tractable expressions for the posterior probability forecast, a normal distribution with mean $\tilde{\mu} = \bm{1}^{\prime} \bm{\Sigma}^{-1} \bm{\mu} / \bm{1}^{\prime} \bm{\Sigma}^{-1} \bm{1}$ and variance $\tilde{\sigma}^{2} = 1/\bm{1}^{\prime} \bm{\Sigma}^{-1} \bm{1}$, where $\bm{1}$ is an $N$-dimensional unit vector, $\bm{\mu}$ is an $N$-dimensional vector of individuals' mean, and $\bm{\Sigma}$ is a known covariance matrix of forecast errors. The mean of the posterior probability forecast is essentially a linear combination of the individuals' means with weights $\bm{1}^{\prime} \bm{\Sigma}^{-1} / \bm{1}^{\prime} \bm{\Sigma} \bm{1}$ identical to the ``optimal'' weights proposed by \citet{Bates1969-yj} and the weights derived from the constrained regression proposed by \citet{Granger1984-jc} (see Section~\ref{sec:linear_comb}), while the latter two approaches do not require normality. The estimation of $\bm{\Sigma}$, therefore, becomes crucial when $\bm{\Sigma}$ is unknown. \citet{Winkler1981-bn} suggested estimating $\bm{\Sigma}$ from data and using an inverted Wishart distribution as a prior for $\bm{\Sigma}$. The procedure is computationally intensive when the number of individual distributions to combine increases; see \citet{Hall2007-lh} for more discussion of the covariance matrix estimation. Following \citet{Winkler1981-bn}, \citet{Palm1992-im} extended the approach to allow for biased individual forecasts, providing a complete solution to the forecast combination problem that takes into account the joint distribution of forecast errors from the individual models.

\citet{Jouini1996-fe} took a different perspective and looked at the likelihood function derived from a copula-based joint distribution, in which dependence among individual distributions is encoded into the copula. The procedure is appealing in the sense of being able to deal with individual forecasts with arbitrary distributions. A recent study by \citet{Wilson2017-ct} gave an expert judgment study to assess the practical significance of the individuals' dependency by comparing the Bayesian combination methods, developed by \citet{Winkler1981-bn} and \citet{Jouini1996-fe}, and common pooling methods.

\subsection{Combinations constructed via integration}

A fully specified Bayesian model is difficult to conceptualize, especially in a setting where biases and miscalibration of individual distributions (and critically, dependencies among them) are time-varying. In this light, the probability forecast combination method of \citet{McAlinn2019-kn} may be helpful. They adapted and extended the basic Bayesian predictive synthesis (BPS) framework developed in agent opinion analysis \citep[see, e.g.,][]{Genest1985-bu,West1992-qy,West1992-gr} to sequential forecasting in time series. In the dynamic extension of BPS model, the posterior probability forecast takes the form
\begin{align*}
  \tilde{f}\left(y_{T+h} | I_{T}, \mathcal{H}_{1: T+h}\right) = \int_{\bm{x}_{T+h}} \alpha\left(y_{T+h} | \bm{x}_{T+h}\right) \prod_{i=1: N} f_{i}\left(x_{T+h,i}|I_{T}\right) d \bm{x}_{T+h},
\end{align*}
where, to use our earlier notation, $\mathcal{H}_{1: T+h}$ denotes the full set of individual probability forecasts available for the decision maker up to forecast origin $T$, $\bm{x}_{T+h} = x_{T+h, 1:N}$ is an $N$-dimensional vector of latent variables at time $T+h$, and $\alpha\left(y_{T+h} | \bm{x}_{T+h}\right)$ is a conditional distribution for $y_{T+h}$ given $\bm{x}_{T+h}$ defining the synthesis function.

Instead of constructing Bayesian combinations by multiplying a likelihood by a prior, the dynamic BPS method follows a subclass of Bayesian updating rules, i.e. updating by integration, in the form of latent factor models. Information about biases, miscalibration, and dependencies among the individual distributions can then be incorporated directly through specification of the synthesis function. Specifically, \citet{McAlinn2019-kn} developed a time-varying (non-convex/nonlinear) combination of probability forecasts by defining a normally distributed synthesis function
\begin{align*}
  \alpha\left(y_{T+h} | \bm{x}_{T+h}\right)=N\left(y_{T+h} | \bm{A}_{T+h}^{\prime} \bm{\theta}_{T+h}, v_{T+h}\right)
\end{align*}
with $\bm{A}_{T+h}=\left(1, \bm{x}_{T+h}^{\prime}\right)^{\prime}$ and $\bm{\theta}_{T+h}=\left(\theta_{T+h, 0}, \theta_{T+h, 1}, \dots, \theta_{T+h, N}\right)^{\prime}$. A dynamic linear model is built to model time evolution of these parameter processes, which is defined as
\begin{align*}
   & y_{T+h}=\bm{A}_{T+h}^{\prime} \bm{\theta}_{T+h}+\nu_{T+h}, \quad \nu_{T+h} \sim N\left(0, v_{T+h}\right),                                                         \\
   & \bm{\theta}_{T+h}=\bm{\theta}_{T+h-1}+\bm{\omega}_{T+h}, \quad \bm{\omega}_{T+h} \sim N\left(\bm{0}, v_{T+h} \bm{W}_{T+h}\right),
\end{align*}
where $\bm{\theta}_{T+h}$ evolves in time according to a normal random walk with innovations variance matrix $v_{T+h} \bm{W}_{T+h}$, and $v_{T+h}$ identifies the residual variance in forecasting $y_{T+h}$ given past information and the set of agent forecast distributions. It is suggested that BPS models be customized specific to the forecast horizon, as a forecasting model may provide different forecast performances at different forecast horizons. MCMC methods are required for this posterior inference and, accordingly, dependencies among agents are involved in sequentially updated estimates of the BPS parameters. Their results of forecasting a quarterly series of inflation rates showed that the proposed dynamic BPS model significantly outperforms benchmark methods, such as the pooling and BMA combination techniques described in Sections~\ref{sec:linear_pooling} and~\ref{sec:bma}. This also held true in a multivariate extension studied by \citet{McAlinn2020-qq}. \citet{McAlinn2019-kn} also showed that the dynamic BPS framework encompasses many existing combination methods, including linear pooling and BMA methods, by specifying different forms of the BPS synthesis function.

\subsection{Quantile forecast combinations}
\label{sec:quantile_comb}

Probabilistic forecasts can also be elicited in the form of quantiles, which are the inverse of the corresponding probability forecasts characterized by the cumulative distribution functions (cdfs). Quantile combinations involve averaging the individuals' quantile functions rather than their inverses as in linear pooling (see Section~\ref{sec:linear_pooling}). In other words, quantile combinations entail horizontally averaging of the individuals' cdfs while linear pooling entails vertically averaging \citep{Lichtendahl2013-rt}.

The standard combination strategy for quantiles is to \textit{allocate identical weights over all quantile levels for each individual model}. For $i=1,\dots,N$, let $F_{i}(y_{T+h}|I_{T})$ denote the individual cdf with corresponding probability density function given by $f_{i}(y_{T+h}|I_{T})$, and let $Q_{T+h|T,i}(\tau) = F_{T+h|T,i}^{-1}(\tau)$ denote the corresponding quantile function. Quantile averaging is then given by
\begin{align}
  \label{eq:quantile_avg}
  \tilde{Q}_{T+h|T}(\tau) = \sum_{i=1}^{N} w_{T+h|T,i} Q_{T+h|T,i}(\tau), \quad 0 < \tau \leq 1,
\end{align}
where the weight $w_{T+h|T,i}\ge0$ such that $\sum_{i=1}^{N}w_{T+h|T,i}=1$. This combination strategy is also referred to as Vincentization \citep{vincent1912-fu}.

Interestingly, unlike linear pooling in Equation~\eqref{eq:linear_pool}, if individual distributions belong to the same location-scale family (such as normal, Logistic, Cauchy, etc.), then quantile averaging yields a combined distribution from the same family, with parameters given by weighted averages of the individuals' parameters \citep{Ratcliff1979-cb,Thomas1980-lp}. Consequently, quantile averaging of normal distributions is always uni-modal (and normal), while linear pooling in general may be multi-modal. Moreover, quantile averaging and linear pooling share the same mean, while quantile averaging tends to be sharper and more confident due to the additional spread driven by the disagreement on the mean in linear pooling.

Is it better to average quantiles (as in quantile averaging) or average probabilities (as in linear pooling)? By restricting themselves to simple averages, \citet{Lichtendahl2013-rt} and \citet{Busetti2017-ox} theoretically and empirically compared the properties of these two combination strategies and suggested that quantile averaging seems overall a preferable and viable approach. \citet{Lichtendahl2013-rt} attributed this, in part, to the fact that the average probability forecast is in general underconfident while the average quantile forecast is always sharper. Even when individual forecasts agree on the location and the average probability forecast is overconfident, the more overconfident average quantile forecast still offers the possibility of forecast improvements due to its shape properties, specifically a higher density in the shoulders and a lower density in the tails. \citet{Busetti2017-ox} reconfirmed the argument and demonstrated that quantile averaging performs better than both linear pooling and logarithmic pooling when combining individual forecast distributions with large biases. Taken together, it may be useful to incorporate quantile averaging and probability averaging so that additional insight may be provided. Accordingly, three simple methods were suggested by \citet{Lichtendahl2013-rt} to blend these two different combination strategies.

Rather than assuming that the entire individual quantile functions are available, one is often provided with a collection of quantiles corresponding to an equidistant dense grid of probabilities $\mathcal{T} \subseteq [0,1]$, leading to a loss of information compared with consideration of the whole distribution. For instance, the $0.05, 0.25, 0.50, 0.75$ and $0.95$ quantiles are often elicited in practice, and another popular choice contains quantiles on all percentiles, i.e. $\mathcal{T} = (0.01, 0.02, \ldots, 0.99)$. Quantile combination in this case turns out to be a special case of the general aggregation rule in Equation~\eqref{eq:quantile_avg}. For each quantile level, equally weighing the quantiles across all individual models is simple and quite robust, yielding improved forecast skill relative to the individual models, and competitive performance relative to a variety of more sophisticated combination strategies \citep[e.g.,][]{Busetti2017-ox,Smyl2019-av,Ray2022-co}. In cases where sufficient past data is available, some effort has been directed toward the determination of combination weights via a cross-validation framework to reflect the past out-of-sample performance of the different models and to improve the utility of combinations. Scoring rules for quantile forecast evaluation can be harnessed for weight construction \citep{Gneiting2007-ij,Grushka-Cockayne2017-eg,Trapero2019-uh}.

One line of research has looked at \textit{tailoring the individual weights for different quantile levels}, i.e., a separate weight is allocated for each individual model and each quantile level, by replacing $w_{T+h|T,i}$ with $w_{T+h|T,i}(\tau)$ in Equation~\eqref{eq:quantile_avg}. For example, individual quantiles can be weighted by the reciprocal of the value of the pinball loss function (also referred to as the quantile loss) \citep{Wang2019-lx,Zhang2020-dm,Browell2020-pa}. This flexible strategy enables the combination to accommodate the fact that individual forecasting models may have varying performances at different quantile levels. However, the number of weights to be learned scales with the number of quantile levels considered, which makes it challenging to achieve forecast improvements. Computationally intensive techniques, such as grid search and linear programming (LP), are applied for weight estimation, which are hardly scalable to large datasets. What's more, the datasets involved in their empirical studies are not large enough to demonstrate the potential benefits of estimating such a large number of weights.

As has been discussed previously in the literature on point forecast combinations, the error in the estimation of optimal weights often exacerbates out-of-sample combined forecasts. The issue is even more problematic when it comes to quantile combinations because it is a much more challenging task to estimate combination weights for a collection of quantiles, especially in the tails of forecast distributions, than merely for point forecasts defined by the means of distributions. For example, \citet{Ray2022-co} failed to empirically demonstrate the utility of weighting different quantiles separately by optimizing the weighted interval score \citep[WIS,][]{Bracher2021-hx} of the corresponding combined result. On the other hand, a promising line of research by \citet{Kim2021-wa}, in the context of quantile regression rather than time series forecasting, produced superior estimates using an aggregation strategy of greater flexibility than previously introduced. Separate weights depend on the features of individual models, and (pairs of) quantile levels. They used a scalable stochastic gradient descent (SGD) algorithm with a monotone operator to solve the weight optimization problem and obviate the need of non-crossing constraints. More recently, \citet{Berrisch2021-cr} introduced a new weighting method that allows the individual forecasters to perform differently over time and within the distribution. This was the first to consider methods that aggregate across quantiles for optimal CRPS-based combinations by using pointwise evaluation across the pinball loss. They also demonstrated the optimal convergence properties, as well as the potential for performance improvements, by considering pointwise optimization of the weight functions.

An independent line of research has looked at model-free combination heuristics, which frequently serve as benchmarks to measure the effectiveness of newly developed combination strategies. From a statistical point of view, these heuristics involve pooling together $Q$ quantiles derived from $N$ individual forecast distributions that are assumed to have the same values of characteristics (e.g., the quantile function at different quantile levels) and using the stacked larger pool to draw more precise estimates of those characteristics without training. \citet{Wang2019-lx} formally introduced na\"{i}ve sorting and median-based sorting methods, in which a total of $N \times Q$ quantiles are stacked and sorted by ascending order to pick the first and median values respectively in consecutive blocks of $N$ forecasts. Naturally, averaging blocks of $N$ forecasts is also another option.

While aggregating quantiles tightly connects with aggregating probability distributions, there has been little theoretical work in this area compared to combinations of probability forecasts which have seen considerable theoretical advances. One exception is \citet{Lichtendahl2013-rt} who looked at the statistical properties of the simple average of probability forecasts and the ability to benefit from averaging. The choice of the combination weights has only been explored empirically, mainly in the context of energy forecasting \citep[e.g.,][]{Wang2019-lx,Browell2020-pa} and epidemiological forecasting \citep[e.g.,][]{Ray2022-co}. Some of these proposals appear practical and beneficial, while some of them appear less useful. Further research is required to explore their utility.

Quantile crossing is a well-known problem caused by the lack of monotonicity in quantile estimates, which may arise when different combination weights are utilized for different quantile levels. Obviously, some model-free heuristics, such as the median, are also included in such cases. Quantile crossing can be avoided by, for example, (i)~integrating the aggregation problems for individual models into one optimization problem subject to more non-crossing constraints \citep[e.g.,][]{Bondell2010-no,Kim2021-wa}, and (ii)~conducting na\"{i}ve rearrangement after all the combined quantiles are obtained \citep[e.g.,][]{Chernozhukov2010-zz,Berrisch2021-cr}. The rearrangement operation, though simple, is frequently recommended in practice since it will never deteriorate the forecasting performance in terms of the pinball loss \citep{Chernozhukov2010-zz}.

Interval forecasts form a crucial special case of quantile forecasts, which makes the preceding combination approaches for quantile forecasts naturally apply to interval forecasts as well. When forming combinations of interval forecasts, attention should be paid to the fact that the combined interval forecasts are not guaranteed to provide target coverage rates \citep{Wallis2005-yf,Timmermann2006-en,Grushka-Cockayne2020-qv}. As a result, when evaluating the combined interval forecasts, proper scoring approaches that take into account both width and coverage are appealing, and can serve as objective functions to determine combination weights; see, e.g., \citet{Gneiting2007-ij} and \citet{Jose2009-lh}.

For interval forecasts, six heuristics have been outlined: (1)~simple average, (2)~median, (3)~envelope, (4)~interior trimming, (5)~exterior trimming, and (6)~probability averaging of endpoints \citep{Park2015-zn,Gaba2017-om}. These six heuristics are virtually free of computational costs and have subsequently been promoted by recent research due to their robustness and benefits in different scenarios for addressing underconfidence/overconfidence; e.g., \citet{Smyl2019-av}, \citet{Petropoulos2020-fp}, and \citet{Grushka-Cockayne2020-qv}. They can easily be extended to address the combinations of quantiles by aggregating individual quantiles in several ways for each quantile level.

Determining combination weights for interval forecasts is evidently easier to implement than quantile forecasts, since one only has to consider two quantiles. For example, by assuming the intervals to be symmetric around the point forecast, \citet{Montero-Manso2020-tq} used the combined point forecast produced by a feature-based meta-learner as the center of the combined interval and generated the radius as a linear combination of the individual radii with the goal of minimizing the MSIS \citep[mean scaled interval score,][]{Gneiting2007-ij} of the interval. The approach achieved the second position in the M4 competition with $100,000$ time series involved. Subsequent work by \citet{Wang2021-un} introduced a feature-based weight determination approach to directly combine lower and upper bounds of individual interval forecasts, leading to significant performance improvements compared to individual forecasts and the simple average.

\section{Conclusions and a look to the future}
\label{sec:conclusion}

Forecasting plays an indispensable role in decision-making, where success depends heavily on the accuracy of the available forecasts. Even with a small increase in accuracy, remarkable gains may be achieved in activities such as management planning and strategy setting \citep{Makridakis1996-cf,Syntetos2009-ho}. In this regard, forecast combinations provide an easy path to improving forecast accuracy by integrating the available information used in individual forecasts.

In this review, our goal has been to show not only how forecast combinations have evolved over time, but also to identify the potential and limitations of various methods, and to highlight the areas needing further research. Forecast combinations can be model-free or model-fitting, linear or nonlinear, static or time-varying, series-specific or cross-learning, and frequentist or Bayesian. The toolbox of combination methods has grown in size and sophistication, each with its own merits. Which combination method to choose depends on several factors such as the form of forecasts (point forecasts, probabilistic forecasts, quantiles, etc.), the quality and size of the model pool, the information available, and the specific forecasting problems. There is no clear consensus on which forecast combination method can be expected to perform best in a specific setting. Based on this review, we summarise some of the current research gaps and potential insights for future research in the following paragraphs.

\textbf{\textit{Continuing to examine simple averaging.}} Over fifty years after \citeapos{Bates1969-yj} pioneering work on forecast combinations, it is amazing that, in empirical studies, simple averaging still repeatedly dominates sophisticated weighted combinations which are theoretically preferred, posing a tough benchmark to beat. Although it is well known that the ``forecast combination puzzle'' stems from the unstable estimates of combination weights, researchers still lack comprehensive quantitative decision guidance on when to choose a simple averaging strategy over more complex strategies. One exception is \citet{Blanc2016-sn} who merely looked at the combination of two individual forecasts and proposed decision rules to decide when to choose simple averaging over the ``optimal'' weights introduced by \citet{Bates1969-yj}. In addition, the examination of simple averaging in the context of probabilistic forecast combinations deserves further attention and development, both theoretical and empirical.

\textbf{\textit{Keeping combinations sophisticatedly simple.}} Forecasting models and forecast combination methods have grown swiftly in both size and sophistication. Nevertheless, empirical results are ambiguous and there is no coherent evidence that complexity systematically improves forecast accuracy; see, e.g., \citet{Green2015-mi} for a review comparing simple and complex forecasting methods. Following \citet{Zellner2001-si}, we suggest the blooming of sophisticatedly simple combinations to balance the tradeoff between the benefits of tailoring weights for different individual models and the instability of learned weights in sophisticated weighting schemes. Additionally, it is strongly recommended that a detailed analysis is required to explore in depth how and why various sophisticated combination strategies work and, thus, provide more insights into which combination method to choose in a particular situation; \citet{Petropoulos2018-fw} provided a good example of this kind of work.

\textbf{\textit{Obtaining statistical inference for the combination forecasts.}} The ``forecast combination puzzle'' revolves primarily around the question of choosing fixed simple weights or random ``optimal'' ones. A related aspect to the puzzle, but somehow different from it, is that the randomness of the combination weights (and in particular the correlation with the forecasts) makes it difficult to perform statistical inference for the weighted combined forecast. A standard error is nontrivial to obtain in most cases, let alone the sampling distribution. Getting the combined forecast is one aspect, what to do with it in a statistical sense is another aspect. Therefore, future studies on the randomness of combination weights and the statistical inference for the combined forecasts would be of interest.

\textbf{\textit{Selecting forecasts to be combined.}} The pool of individual forecasts lays the foundation for the utility of forecast combinations. These forecasts may come from statistical or machine learning models, and may be based on observed data or be elicited from experts. Empirical evidence suggests that the future lies in the combination of statistical and machine learning generated forecasts, as well as the incorporation of human judgment \citep{Petropoulos2018-ad,Makridakis2020-hu,Petropoulos2021-ft}. Given a large number of available forecasts, selecting a subset of combinations becomes particularly pivotal for the purpose of improving forecast skill and reducing computational costs. A variety of crucial issues in the selection of forecasts have to be addressed, such as accuracy, robustness, diversity, and calibration when considering probabilistic forecasts \citep{Lichtendahl2020-ut,Wang2022-tr}. However, most of the existing algorithms perform ad hoc selections and lack statistical rationale \citep{Kourentzes2019-na}. Therefore, further attention should be paid to the development of empirical guidelines and quantitative metrics to help forecasters in selecting forecasts prior to combinations. Since a zero weight in the past does not indicate a zero weight in the future, time-varying subset selection for forecast combination would also be an interesting research direction.

\textbf{\textit{Advancing the theory of nonlinear combinations.}} While the benefits of linear combinations of multiple forecasts are well appreciated in the forecasting literature, less attention has been paid to nonlinear combination schemes to model nonlinear dependencies among individual forecasts possibly due to the lack of theoretical foundations and poor records of success; see \citet{Timmermann2006-en} for a brief review of the related literature. Nonlinearities are currently addressed with the use of neural networks or an additional nonlinear term in the combination equation. However, the limited evidence on the benefits of involving nonlinear combinations is mostly derived from only a few time series and is not entirely unconvincing. As a consequence, we expect more theoretical and empirical work in this area in the near future.

\textbf{\textit{Focusing more on probabilistic forecast combinations.}} In probabilistic forecast combinations, linear pooling and quantile averaging suggest two different ways of thinking --- linear pooling entails vertically averaging the individuals' cdfs while quantile aggregation entails horizontally averaging. Accordingly, their combined forecasts hold different properties and benefit differently from the combination. For example, the shape-preserving property of quantile averaging may be appealing in certain settings \citep{Lichtendahl2013-rt}. Over the past decade, linear pooling has attracted considerable attention, achieving appreciable advancements both theoretically and empirically. Quantile averaging, however, has not received much attention, especially in the theoretical realm. Furthermore, when tailoring combination weights for different quantile levels, the instability of the estimated weights is especially problematic since a lot of parameters have to be estimated. This issue is likely to harm the calibration and sharpness of the out-of-sample combined forecasts, making quantile averaging a challenging task. Taken together, we expect combinations of quantiles to be an important area of research in the future.

\textbf{\textit{Discussing if, how, and when it is helpful to interpret combination weights.}} In probability forecast combinations, some combination approaches have the property that poorly performing forecasts will almost surely be rejected in favor of the best one as the sample size tends to infinity. For example, BMA reduces to model selection for a large sample size, with the best model receiving a weight very close to one. See Section~\ref{sec:bma} for more detailed discussions. However, it is sometimes found that individually ``bad'' forecasts may be still helpful in combinations~\citep[e.g.,][]{Geweke2011-xk}. In this case, one does not want to zero-weight these bad forecasts (in the limit, as the sample size goes to infinity). This relates to the question of if, how, and when it is helpful to interpret combination weights, another future research direction worth exploring.

\textbf{\textit{Taking account of correlations among individual forecasts.}} Some sort of correlations among individual forecasts are expected as they are likely to share the same data, overlapping information, similar forecasting models, and a common training process. Such correlations can be critical and have a serious impact on the utility of forecast combinations \citep{De_Menezes2000-vd}. An extensive body of literature on point forecast combinations has attempted to account for correlations in terms of weight estimation, despite the fact that these correlations can be poorly estimated. Despite the existence of such correlations, the literature on probabilistic forecast combinations has paid scant attention to addressing them; they are primarily addressed from a Bayesian perspective \citep[e.g.,][]{Winkler1981-bn,McAlinn2019-kn}. Therefore, another interesting path for further research would be to take more into account of correlations among individual forecasts in weighting schemes for probabilistic forecast combinations.

\textbf{\textit{Cross-learning and feature engineering.}} Instead of combinations in a series-by-series fashion, numerous studies have confirmed the beneficial usage of information from multiple series to study common patterns among series, thereby facilitating the determination of combination weights and exploiting the benefits of cross-learning. The evidence of the potential of cross-learning has largely come from competitions \citep[e.g.,][]{Makridakis2020-hu,Makridakis2020-fn} and empirical studies \citep[e.g.,][]{Ma2021-np}. Moreover, access to feature engineering can lead to improved forecasting performance, providing valuable information for forecast combinations in a cross-learning fashion \citep{Montero-Manso2020-tq,Kang2021-ol}. In this regard, we believe that further research needs to be done on feature engineering for time series data to unlock the potential of cross-learning.

\textbf{\textit{Encouraging researchers to contribute open-source software and datasets.}} In this paper, we list some open-source packages linking to the developed approaches for forecast combinations (e.g., \pkg{fable}, \pkg{ForecastComb}, and \pkg{forecastHybrid} packages for \proglang{R}), time series features (e.g., \pkg{feasts} and \pkg{tsfeatures} packages for \proglang{R} and \pkg{tsfresh} and \pkg{Kats} packages for \proglang{Python}), and time series generations (e.g., \pkg{forecast} and \pkg{gratis} packages for \proglang{R}). We emphasize that open-source research software is a pathway to impact. Recent decades have witnessed a dramatically accelerating pace in the advancements in computing. As a consequence, it is time to promote the idea of researchers producing open-source software that provides evidence and support behind all the statements that are made. Publicly releasing new software benefits researchers and end users. It reduces research costs, allows for quick implementation, helps people modify the existing software and adapts it to other research ideas. We also encourage researchers to contribute open-source datasets because of the benefits of investigating and comparing the performance of newly developed methods; see, e.g., \citet{Godahewa2021-mo,Godahewa2021-wb} for a time series forecasting archive containing $20$ publicly available time series datasets from different domains.

In this paper, we take the multiple forecasts to be combined essentially as given and limit ourselves to combinations of forecasts derived from separate models for a given series. These separate models can be identified with different model forms and/or the same model form with different parameters. However, we highlight that there are other types of forecast combinations in the forecasting literature. For example, one type of approach involves constructing replicas of the original time series through various manipulations of local curvatures, frequency transformation or bootstrapping, and subsequently multiple forecasts are produced to form the final combined forecasts, leading to a wide variety of approaches such as the theta method \citep{Assimakopoulos2000-cc}, temporal aggregation \citep[e.g.,][]{Kourentzes2014-wz,Kourentzes2016-qq,Kourentzes2017-xe}, bagging \citep[e.g.,][]{Bergmeir2016-ae,Petropoulos2018-fw}, and structural combination \citep[e.g.,][]{Rendon-Sanchez2019-qm}. Another approach involves forming a hierarchical structure using multiple time series that are structurally connected based on geographic or logical reasons and reconciling multiple forecasts across the hierarchy, leading to various hierarchical aggregation methods \citep[e.g.,][]{Hyndman2011-sd,Wickramasuriya2019-fc,Taieb2021-tc,Hollyman2021-tn}.

\section*{Acknowledgments}

We thank Adrian Raftery, Casey Lichtendahl, Yael Grushka-Cockayne, Fotios Petropoulos and other experts in this area for providing very helpful feedback on an earlier version of this paper. We thank the editors and two anonymous reviewers for their helpful comments and suggestions that improved the paper.

\newpage
\printbibliography

\end{document}